\documentclass[
  reprint,
  superscriptaddress,
  amsmath,
  amssymb,
  aps,
  longbibliography
]{revtex4-2}

\usepackage[utf8]{inputenc}
\usepackage[lining,semibold]{libertine}
\usepackage{amsthm}
\usepackage[libertine,cmintegrals,bigdelims,vvarbb]{newtxmath}

\usepackage{amsmath}
\usepackage{amsfonts}
\usepackage{bm}
\usepackage{mathrsfs}

\usepackage{graphicx}
\usepackage{dcolumn}
\usepackage{booktabs}
\usepackage{siunitx}

\usepackage{xcolor}
\definecolor{webgreen}{rgb}{0,.5,0}
\definecolor{webbrown}{rgb}{.6,0,0}
\definecolor{RoyalBlue}{rgb}{0.0,0.14,0.4}

\usepackage{hyperref}
\hypersetup{
  colorlinks=true,
  linktocpage=true,
  breaklinks=true,
  pdfstartpage=1,
  pdfstartview=FitV,
  pdfpagemode=UseOutlines,
  plainpages=false,
  bookmarksnumbered=true,
  bookmarksopen=true,
  bookmarksopenlevel=1,
  hypertexnames=true,
  pdfhighlight=/O,
  urlcolor=webbrown,
  linkcolor=RoyalBlue,
  citecolor=webgreen,
  pdfauthor={Timur Aslyamov},
  pdfcreator={pdfLaTeX},
  pdfproducer={LaTeX REVTeX}
}

\usepackage[capitalize]{cleveref}

\crefname{appendix}{Appendix}{Appendices}
\Crefname{appendix}{Appendix}{Appendices}

\makeatletter
\def\maketag@@@#1{\hbox{\m@th\normalfont\normalsize#1}}
\makeatother

\newcommand{\llangle}{\langle\!\langle}
\newcommand{\rrangle}{\rangle\!\rangle}

\newcommand{\blue}[1]{\textcolor{black}{#1}}

\begin{document}
\title{Dynamical Fluctuation-Response Relations}

\author{Timur Aslyamov}
\email{timur.aslyamov@uni.lu}
\affiliation{Complex Systems and Statistical Mechanics, Department of Physics and Materials Science, University of Luxembourg, 30 Avenue des Hauts-Fourneaux, L-4362 Esch-sur-Alzette, Luxembourg}

\author{Massimiliano Esposito}
\email{massimiliano.esposito@uni.lu}
\affiliation{Complex Systems and Statistical Mechanics, Department of Physics and Materials Science, University of Luxembourg, 30 Avenue des Hauts-Fourneaux, L-4362 Esch-sur-Alzette, Luxembourg}

\date{\today}


\begin{abstract}
We derive exact dynamical fluctuation-response relations (FRRs) for time-integrated observables of any nonautonomous Markov jump process. The finite-time covariance splits into an initial variability and an integral of response kernels along the driven dynamics. The identity sharpens the dynamical response thermodynamic and kinetic uncertainty relations and fluctuation-response inequalities (FRIs).
\blue{For autonomous processes, dynamical FRRs yield two complementary frequency-domain relations. 
Steady-state FRRs are recovered in the long time limit, along with the fluctuation-dissipation theorem and Onsager reciprocity for detailed-balance dynamics.
The importance of initial variability for slowly relaxing systems is also highlighted.}
\end{abstract}

\maketitle

\textit{Introduction---}At equilibrium, the fluctuation-dissipation theorem (FDT) identifies linear response coefficients with spontaneous fluctuations~\cite{kubo1966fluctuation,kubo2012statistical,stratonovich2012nonlinear,marconi2008fluctuation,forastiere2022linear}. 
Away from equilibrium, response theory remains possible but usually takes a different form: the response of an observable is written as a correlation with an auxiliary observable conjugated to the perturbation~\cite{agarwal1972fluctuation,speck2006restoring,chetrite2008fluctuation,baiesi2009fluctuations,baiesi2009nonequilibrium,seifert2010fluctuation,maes2020response,prost2009generalized,zheng2025nonlinear,stutzer2025stochastic}. 
Such identities are useful but do not directly express the variance or covariance of the observable whose response is measured. 
This distinction is important in stochastic thermodynamics, where fluctuations of currents and state observables provide experimentally accessible information about dissipation, dynamical activity, kinetic bottlenecks, and thermodynamic efficiency~\cite{barato2015thermodynamic,gingrich2016dissipation,pietzonka2016universal,pietzonka2017finite,horowitz2017proof,falasco2020unifying,horowitz2020thermodynamic,vu2020entropy,ohga2023thermodynamic,van2023thermodynamic,di2018kinetic,shiraishi2021optimal,dechant2021improving,kwon2024unified,dieball2023direct}.

A different connection between fluctuations and responses has recently emerged for nonequilibrium steady states. 
Exact fluctuation-response relations (FRRs) for Markov jump processes express covariances of time-integrated state and current observables directly in terms of responses to microscopic perturbations of transition rates, without introducing a conjugate auxiliary observable~\cite{aslyamov2025frr,ptaszynski2024frr,ptaszynski2025frr-mixed}. 
These relations provide the structural basis for response thermodynamic uncertainty relations (R-TURs) and response kinetic uncertainty relations (R-KURs), which bound dissipation or activity by response-to-fluctuation ratios~\cite{zheng2024information,liu2024dynamical,ptaszynski2024dissipation,kwon2024fluctuation,bao2024nonequilibrium,van2024fundamental,liu2025response,monnai2026entropic,zheng2026thermodynamic,gu2026spectral}. 
They also survive in macroscopic weak-noise limits and in continuum limits under diffusive scaling~\cite{dechant2025finite,aslyamov2026macroscopic,aslyamov2025excess,chun2026fluctuation}. 
However, their exact form has so far been restricted to steady states. 
This leaves out relaxation from arbitrary initial states and systems driven by time-dependent protocols, even though dynamical thermodynamic uncertainty relations (TURs) and fluctuation-response inequalities (FRIs) are known to hold in these regimes~\cite{dechant2020fluctuation,koyuk2020thermodynamic,liu2020thermodynamic,kwon2024fluctuation,liu2024dynamical,van2024fundamental,kwon2025unified,chun2026fluctuation}.

In this Letter, we derive exact dynamical FRRs for arbitrary time-integrated state and \blue{transition} observables of nonautonomous Markov jump processes. 
The central result is a finite-time covariance identity with two contributions. 
The first is an integral over products of dynamical response kernels to delta-like perturbations of local transition parameters. 
The second is an initial-variability term measuring how much the conditional mean of the observable depends on the initial state. 
For variances, this term is nonnegative. 
The initial variability term is zero for deterministic initial preparation.
\blue{It also vanishes over long times for systems that eventually forget how they were prepared.
For systems with slow relaxation (small spectral gaps), it can be long-lived and even dominant over large transients.}
In the infinite time limit, our dynamical FRR reduces to the known steady-state FRRs.

\begin{figure}
    \centering
    \includegraphics[width=\linewidth]{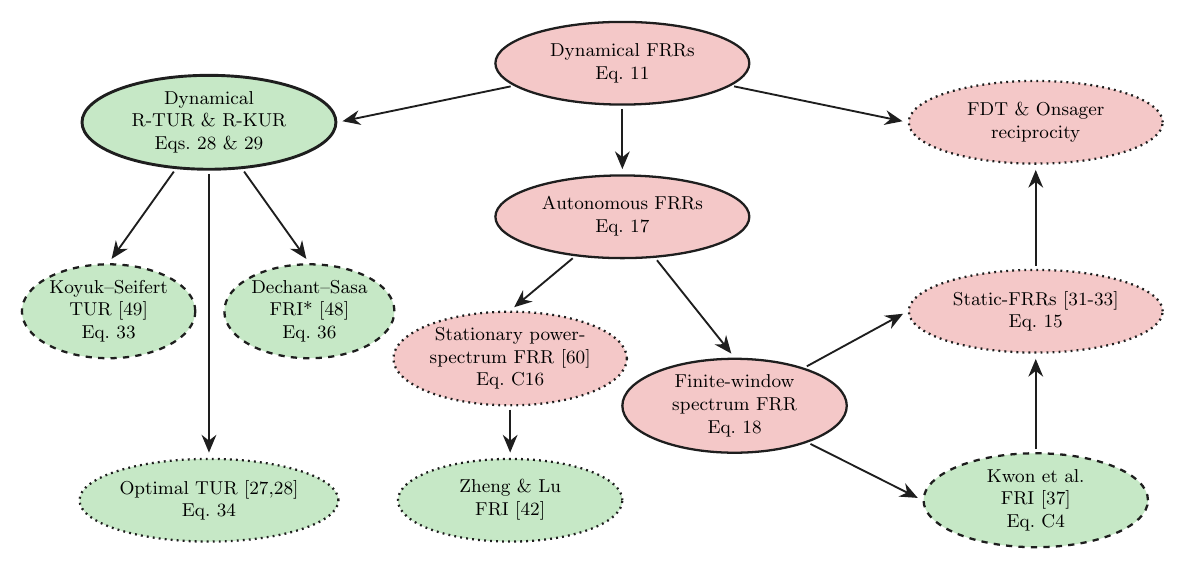}
    \caption{Hierarchy of the main results that follow from the dynamical FRRs. Red nodes denote exact identities; green nodes denote bounds. Solid outlines indicate results derived in this Letter, while dashed outlines indicate previously known results refined by the initial-variability term, and dotted outlines indicate known results derived without changes. The Dechant--Sasa FRI is marked by $*$ because here only its linear-response Markov-jump version is derived.}
    \label{fig:hierarchy}
\end{figure}

Importantly, the dynamical FRR unifies a broad class of central results and bounds developed in stochastic thermodynamics over the last decade, while sharpening their finite-time forms.
It yields dynamical R-TURs, improves finite-time TURs~\cite{koyuk2020thermodynamic,liu2020thermodynamic}, R-KURs~\cite{zheng2024information,liu2024dynamical,van2024fundamental}, and Dechant--Sasa FRI~\cite{dechant2020fluctuation} by retaining the positive initial-variability contribution, and reduces, once memory of the preparation is lost, to steady-state FRRs~\cite{aslyamov2025frr,ptaszynski2024frr,ptaszynski2025frr-mixed} and optimal steady-state TUR bounds~\cite{shiraishi2021optimal,dechant2021improving}. This hierarchy is summarized in \cref{fig:hierarchy}.

\textit{Setup---}We consider a Markov jump process on the discrete states $n\in\{1,\dots,N\}$. 
The probability that the system occupies state $n$ at time $t$ is denoted by $\bm{p}(t)=(p_1, \dots, p_N)^\intercal$ and satisfies the master equation $d_t \bm{p}(t) = W(t) \cdot \bm{p}(t)$,
where $\sum_n p_n = 1$ and where $\bm{p}(0)$ is assumed to be known. 
\blue{For a nonautonomous process, the rate matrix $W(t)$ explicitly depends on time through a prescribed protocol. For an autonomous irreducible one it does not, $W(t)=W$, and any initial distribution will eventually relax to the stationary one.
The process can always be represented on an oriented graph where nodes denote the states $n$ and edges describe transitions (jumps) between states. The forward and reverse orientations of an edge are denoted by $e$ and $-e$, respectively.
The off-diagonal elements of $W$ are given by $W_{nm}(t)=\sum_{e}W_{e}(t)\delta_{s(e)m}\delta_{t(e)n}$, 
where $\sum_{e}f_{\pm e} \equiv \sum_{e>0}(f_{e} + f_{-e})$ denotes summation over all edges in both orientations. Here, $s(e)$ and $t(e)$ are the source and target of edge $e$, respectively, with $s(e)=t(-e)$ and $t(e)=s(-e)$. The diagonal elements of $W$ are $W_{nn}=-\sum_{m \neq n}W_{mn}$.
}
\blue{The graph incidence matrix is~$\bm \Delta_{n e} = \delta_{n t(e)}-\delta_{n s(e)}$ and} 
the incidence vector is $\bm{\Delta}_{e}=(\dots,\bm \Delta_{n e},\dots)^\intercal$ such that  $\bm{\Delta}_e \cdot \bm f = f_{t(e)} - f_{s(e)}$ and $\bm \Delta_{e} = - \bm \Delta_{-e}$.

The stochastic trajectory of the system from time $0$ to $T$ is denoted by $\omega_{0:T} = \{n_t\}_{0\le t\le T}$. A generic integrated observable along it reads
\begin{align}
\label{eq:observ-def}
    \mathcal{Q}[\omega_{0:T}] &= \underbrace{\int_{0}^{T} d\tau o_{n_\tau}(\tau)}_{\mathcal{O}[\omega_{0:T}]} + \underbrace{\int_{0}^{T} \sum_{e} x_{e}(\tau) dk_{e}(\tau)}_{\mathcal{J}[\omega_{0:T}]}\,.
\end{align}
Here $\mathcal{O}[\omega_{0:T}]$ is a state observable part, with $o_{n_\tau}(\tau)$ a function defined on the states of the trajectory, while $\mathcal{J}[\omega_{0:T}] $ is a \blue{transition} observable part, with \blue{oriented} edge weights $x_{\pm e}(\tau)$ (which may explicitly depend on time) and 
\blue{$dk_{e}(\tau)=1$ (respectively, $dk_{-e}(\tau)=1$)} if a transition \blue{$e$ (respectively, $-e$)} occurs along the trajectory in $[\tau,\tau+d\tau]$; the corresponding counter is zero otherwise. 
\blue{Antisymmetric weights, $x_{e} = -x_{-e}$, define a current counter $dk_{e}-dk_{-e}$, whereas symmetric weights, $x_{e} = x_{-e}$, define an activity counter $dk_{e}+dk_{-e}$.}
The average current and activity read $I_e(t) = j_{e} - j_{-e}$ and $a_e(t)= j_{e} + j_{-e}$, respectively, where $j_{e}\equiv W_{e}(t)p_{s(e)}(t)$.
The average currents define the entropy production rate (EPR) as $\sigma(t)=\sum_{e>0} \sigma_e(t)$ with $\sigma_e (t) = I_e(t)\ln [j_{e}(t)/j_{-e}(t)]$, while a combination of currents and activities defines the pseudo-EPR $\sigma^\text{ps}(t) = \sum_{e>0} \sigma_e^\text{ps}(t)$, where $\sigma_e^\text{ps}(t)=I_e(t)^2/a_e(t)$, such that $\sigma(t) \geq 2\sigma^\text{ps}(t)$. 
\blue{
The pseudo-EPR is a nonnegative measure of irreversibility that vanishes at equilibrium. Close to equilibrium, where $|I_e|/a_e\ll 1$, it coincides to leading order with half the EPR, since
$\sigma_e/2=I_e\operatorname{arctanh}(I_e/a_e)=\sigma_e^\text{ps}+\mathcal{O}(I_e^4/a_e^3)$.
Beyond the near-equilibrium regime, the pseudo-EPR plays a central role in sharpened and optimal TUR-type bounds~\cite{dechant2021improving,shiraishi2021optimal,van2022unified,dechant2022minimum}.
}
 
Introducing the trajectory probability density $\mathbb{P}[\omega_{0:T}]$, we define the mean of \cref{eq:observ-def} as
\begin{align}
\label{eq:mean-def}
    Q(0, T) \equiv \sum_{\omega} \mathcal{Q}[\omega_{0:T}] \mathbb{P}[\omega_{0:T}]\,.
\end{align}
\blue{Here and below, $\sum_{\omega}$ denotes the full path-space integral,
including sums over the initial state, the number and sequence of jumps,
and integrations over the ordered jump times.}
It has a time-integral form \blue{$Q(0, T) = \int_0^T dt \dot Q(t)$}  and can be computed as
\begin{align}
Q(0, T) &= \underbrace{\int_{0}^{T} d\tau \sum_n o_n(\tau) p_n(\tau)}_{O(0,T)} + \underbrace{\int_{0}^{T} d\tau\sum_{e} x_{e}(\tau) j_{e}(\tau)}_{J(0, T)}\,,
\end{align}
where $O(0, T)$ and $J(0, T)$ are the means of the state and \blue{transition} observables, respectively. 

Using the conditional probability density $\mathbb{P}[\omega_{\tau:T}|n_{\tau} = n]$ for trajectories starting at time $\tau$ from state $n$, we introduce the conditional mean:
\begin{align}
\label{eq:mean-cond-def}
    Q_n(\tau,T) \equiv \sum_{\omega} \mathcal{Q}[\omega_{\tau:T}] \mathbb{P}[\omega_{\tau:T}|n_{\tau} = n]\,,
\end{align}
which defines $\bm{Q}(\tau, T) \equiv  (Q_1(\tau, T), \dots, Q_N(\tau, T))^\intercal$.
Naturally, $Q(\tau, T)=\sum_n Q_n(\tau,T) p_n(\tau)$. 
Let us note that a useful tool for computing $\bm{Q}(\tau, T)$ is solving the backward Kolmogorov equation~\cite{van2006total}
(derived in \cref{sec:DFRR}),
\begin{align}
\label{eq:Kolmogorov}
    -\partial_{\tau} \bm{Q}(\tau, T) =  W^\intercal(\tau)\cdot \bm{Q}(\tau, T) + \blue{\bm \nabla_{p}\dot Q(\tau)}\,,
\end{align}
with terminal condition $\bm{Q}(T, T) = 0$, and where
\blue{$\bm \nabla_{p} \equiv (\dots, \partial_{p_n}, \dots)^\intercal$ and 
$\partial_{p_n}\dot Q(\tau) = o_n(\tau)  + \sum_{e}x_{e}(\tau) W_{e}(\tau) \delta_{n, s(e)}$}.
\Cref{eq:Kolmogorov} can be seen as a dynamical extension of the Poisson equation \cite{khodabandehlou2024poisson}.

\textit{Dynamical FRRs---}The first central object is the scaled covariance between 
$\mathcal{Q}[\omega_{0:T}]$ and $\mathcal{Q}'[\omega_{0:T}]$ corresponding to the pairs $\{x_e, o_n\}$ and $\{x'_e, o'_n\}$: 
\begin{align}
\label{eq:C-def}
    \langle\!\langle \mathcal{Q}, \mathcal{Q}' \rangle\!\rangle \equiv \frac{1}{T}\sum_{\omega}\Delta \mathcal{Q}[\omega_{0:T}]\Delta \mathcal{Q}'[\omega_{0:T}]\mathbb{P}[\omega_{0:T}]\,,
\end{align}
where $\Delta \mathcal{Q}[\omega_{0:T}] \equiv \mathcal{Q}[\omega_{0:T}] - Q(0, T)$.

The second central object is the response kernel to a perturbation \blue{$\delta\lambda(t)$ on an arbitrary protocol $\lambda(t)$ driving the rates $W(t;\lambda(t))$ which define the unperturbed reference dynamics.}
\blue{The response kernel is defined as a functional derivative evaluated along unperturbed reference dynamics:}
\begin{align}
\label{eq:kernel-def}
    R_{\lambda}(\tau,T)\equiv \frac{\delta Q(0,T)}{\delta \lambda(\tau)}
    = \frac{\delta Q(\tau,T)}{\delta \lambda(\tau)}\;,
\end{align}
which describes the linear response of the observable $\delta Q(0,T) \equiv Q^{\mathrm{pert}}(0,T)-Q(0,T)$ to the time-dependent perturbation
$\lambda(t) \mapsto \lambda(t)+\delta \lambda(t)$:
\begin{align}
\label{eq:dQ}
    \delta Q(0,T)
    = \int_0^T d\tau R_{\lambda}(\tau,T)\,\delta \lambda(\tau) + \mathcal O[(\delta\lambda)^2]\,.
\end{align}
We note that for a delta perturbation $\delta \lambda(t)= h \delta(t-\tau)$, $\delta Q(0,T)= h R_{\lambda}(\tau,T) + \mathcal O(h^2)$,
while for step perturbation $\delta\lambda(t)=h\Theta(t)$, $\delta Q(0,T)=h T\overline{R}_\lambda+ \mathcal O(h^2)$, where $\overline{f} \equiv T^{-1}\int_0^T d\tau f(\tau)$.

We show in \cref{sec:DFRR} that the response kernel can be rewritten in a convenient form 
\begin{subequations}
\label{eq:kernel-phi}
    \begin{align}
\label{eq:kernel-general}
    R_{\lambda}(\tau, T) &= \sum_{e} \varphi_{e}(\tau, T) \blue{j_{e}(\tau) \frac{\partial \ln W_{e}(\tau)}{\partial \lambda(\tau)}}\,,\\
\label{eq:varphi}
    \varphi_{e}(\tau, T) &\equiv x_{e}(\tau) + \bm{\Delta}_e\cdot\bm{Q}(\tau, T)\,,
\end{align}
\end{subequations}
where only derivatives in $\lambda$ of the rate matrix now appear. 
For single \blue{oriented} edge perturbations $\lambda_{e}(t)$ such that $W_{e}\equiv W_{e}(t; \lambda_{e}(t))$, \cref{eq:kernel-general} simplifies to:
\begin{align}
\label{eq:kernel}
    R_{\lambda_{e}}(\tau, T) = \varphi_{e}(\tau, T) j_{e}(\tau) \frac{\partial \ln W_{e}(\tau)}{\partial \lambda_{e}(\tau)}\,.
\end{align}
In our derivations, we assume constant observable weights $\partial_\lambda x_{e}(\tau)=\partial_\lambda o_n(\tau)=0$. For general weights, the same formulas hold after replacing $R_\lambda(\tau,T)$ by 
$R_\lambda(\tau,T)-\sum_n p_n(\tau)\partial_\lambda o_n(\tau)-\sum_e j_e(\tau)\partial_\lambda x_e(\tau)$.

Our central result is the dynamical Fluctuation-Response Relation (FRR) relating covariance and responses: 
\begin{align}
\label{eq:FRR}
  \langle\!\langle \mathcal{Q}, \mathcal{Q}' \rangle\!\rangle &=\langle\!\langle\mathcal{Q}, \mathcal{Q}' \rangle\!\rangle_{0}+ \int_0^T \frac{dt}{T} \sum_{e} j_{e}(t)\varphi_{e}(t, T)\varphi'_{e}(t, T)\nonumber\\
  &\hspace{-1cm}=\langle\!\langle\mathcal{Q}, \mathcal{Q}' \rangle\!\rangle_{0}+ \int_0^T \frac{dt}{T} \sum_{e} \frac{R_{\lambda_{e}}(t, T)R'_{\lambda_{e}}(t, T)}{j_{e}(t)[\partial_{\lambda_{e}}\ln W_{e}(t)]^2}\,,
\end{align}
which we derive in \cref{sec:DFRR} using the finite-time cumulant-generating functions~\cite{lebowitz1999gallavotti,esposito2007entropy}.
For variance, it reduces to 
\begin{align}
\label{eq:var-FRR}
    \langle\!\langle \mathcal{Q}^2 \rangle\!\rangle =\langle\!\langle \mathcal{Q}^2 \rangle\!\rangle_{0} + \frac{1}{T}\int_0^T dt \sum_e \frac{R^2_{\lambda_e}(t, T)}{j_e(t)[\partial_{\lambda_e} \ln W_e(t)]^2} \,,
\end{align}
where $\langle\!\langle \mathcal{Q}^2 \rangle\!\rangle \equiv \langle\!\langle \mathcal{Q}, \mathcal{Q} \rangle\!\rangle$.
The FRR relates covariance and responses via a third central quantity, which we call \textit{initial variability}:
\begin{align}
\label{eq:Omega}\langle\!\langle\mathcal{Q}, \mathcal{Q}' \rangle\!\rangle_{0} 
    &\equiv\frac{1}{T}\sum_{n}\Delta Q_{n}(0, T)\Delta Q'_{n}(0, T) p_n(0) \,,
\end{align}
where $\Delta Q_{n}(t, T) \equiv Q_{n}(t, T) - Q(t, T)$ are the excess observables measuring the difference between a conditional mean over the interval $[t, T]$ given the state at time $t$ and its unconditional average. \blue{Physically, $\llangle\mathcal{Q}, \mathcal{Q}'\rrangle_{0}$ separates the preparation-induced variability from the dynamical fluctuations generated after $t=0$.}
It vanishes when the initial state lacks variability, i.e. when $p_n(0)=\delta_{nk}$ and thus $\Delta Q_k(0,T)=0$.

We emphasize that the FRR is exact. The response kernels are functional derivatives evaluated along the unperturbed dynamics. \blue{They describe the linear response to protocol perturbations, but the FRR itself does not involve any linearization.}

\blue{\textit{Autonomous relaxation---}We now consider autonomous dynamics (time-independent $W$, $\bm x$, and $\bm o$) with a generic nonstationary initial distribution $\bm p(0)\neq\bm\pi$ that eventually relaxes toward the steady state $\bm\pi$.} 
\blue{Under these assumptions, \cref{eq:Kolmogorov} gives $\bm Q(t,T)=\int_t^T ds\,e^{W^\intercal(s-t)}\bm\nabla_p\dot Q$. Let $\bm\ell^{(k)}$, $\bm r^{(k)}$, and $\lambda_k$ denote the left eigenvectors, right eigenvectors, and eigenvalues of $W$, respectively, normalized as $\bm r^{(k)}\cdot\bm\ell^{(k')}=\delta_{kk'}$. 
Assuming that $W$ is diagonalizable, with $\lambda_0=0$ and spectral gap $\gamma=-\max_{k\ne0}\operatorname{Re}\lambda_k$, the decomposition $e^{W^\intercal s}=\bm 1\bm\pi^\intercal+\sum_{k=1}^{N-1}e^{\lambda_k s}\bm\ell^{(k)}\bm r^{(k)\intercal}$ yields
\begin{align}
\label{eq:excess-relax}
    \Delta Q_n(0,T)
    =\sum_{k=1}^{N-1}\Delta\ell_n^{(k)}
    (\bm r^{(k)}\cdot\bm\nabla_p\dot Q)
    \frac{e^{\lambda_kT}-1}{\lambda_k}\,,
\end{align}
where $\Delta\ell_n^{(k)}\equiv\ell_n^{(k)}-\bm\ell^{(k)}\cdot\bm p(0)$. For $T \to 0$, the excess observables \eqref{eq:excess-relax} scale as $\Delta Q_n(0,T)\propto T$, implying that $\llangle \mathcal{Q}^2 \rrangle_0 \propto T$ [\cref{eq:Omega}]. In the long-time limit $T\to \infty$, the excess observables \eqref{eq:excess-relax} saturate at their steady-state values, and, as a result, the factor $1/T$ in \cref{eq:Omega} washes out the effects of the initial variability, $\lim_{T\to\infty}\llangle\mathcal Q^2\rrangle_0=0$. Therefore, the initial variability reaches its maximum at intermediate times. 
}
\blue{Also, when $T\to\infty$, the response kernel in the temporal bulk approaches the static steady-state response, $R_{\lambda_e}(\tau,T)\to R_{\lambda_e}^\text{ss}$, where $R_{\lambda_e}^\text{ss}=\sum_n o_n\partial_{\lambda_e}\pi_n+\sum_{e'}x_{e'}\partial_{\lambda_e}j_{e'}^\text{ss}$ with $j_e^\text{ss} = W_e \pi_{s(e)}$. 
Thus, \cref{eq:var-FRR} reduces to a unified form of the previously derived static FRR~\cite{aslyamov2025frr,ptaszynski2024frr,ptaszynski2025frr-mixed}: 
\begin{equation}
\lim_{T\to\infty}\llangle\mathcal Q^2\rrangle=\sum_e \frac{(R_{\lambda_e}^\text{ss})^2}{j_e^\text{ss}(\partial_{\lambda_e}\ln W_e)^2}\;.
\end{equation}
}

\blue{
We now seek a regime where the initial variability may dominate \cref{eq:var-FRR}.
If the relaxation is dominated by a single real slow mode, $\lambda_1=-\gamma$, the initial variability is
\begin{align}
\label{eq:variability-relax}
    \llangle\mathcal Q^2\rrangle_0
    \approx A_Q(\gamma)\frac{(1-e^{-\gamma T})^2}{T\gamma^2}\,,
\end{align}
where $A_Q(\gamma)\equiv(\bm r^{(1)}\cdot\bm\nabla_p\dot Q)^2\sum_n p_n(0)(\Delta\ell_n^{(1)})^2$. As the gap closes, suppose that $A_Q(\gamma)\sim\gamma^\alpha$ and that the slow mode remains dominant in \cref{eq:excess-relax}.
%
In the joint slow-relaxation limit $\gamma\to0$ with $\gamma T=\mathcal O(1)$, it scales as $\llangle\mathcal Q^2\rrangle_0\sim\gamma^{\alpha-1}$. Thus, for $\alpha<1$, the initial variability remains appreciable over the slowest relaxation time, opening a regime in which it can dominate the dynamically generated contribution. The long-lived scaling holds also for oscillating slow mode $\lambda_{1,2} = -\gamma \pm i\omega$ because the frequency is bounded by $|\omega|\leq \gamma \cot(\pi/N)$~\cite{dmitriev1946characteristic,ohga2023thermodynamic}; see details in \cref{sec:auto}.
}

\begin{figure}
    \centering
    \includegraphics[width=\linewidth]{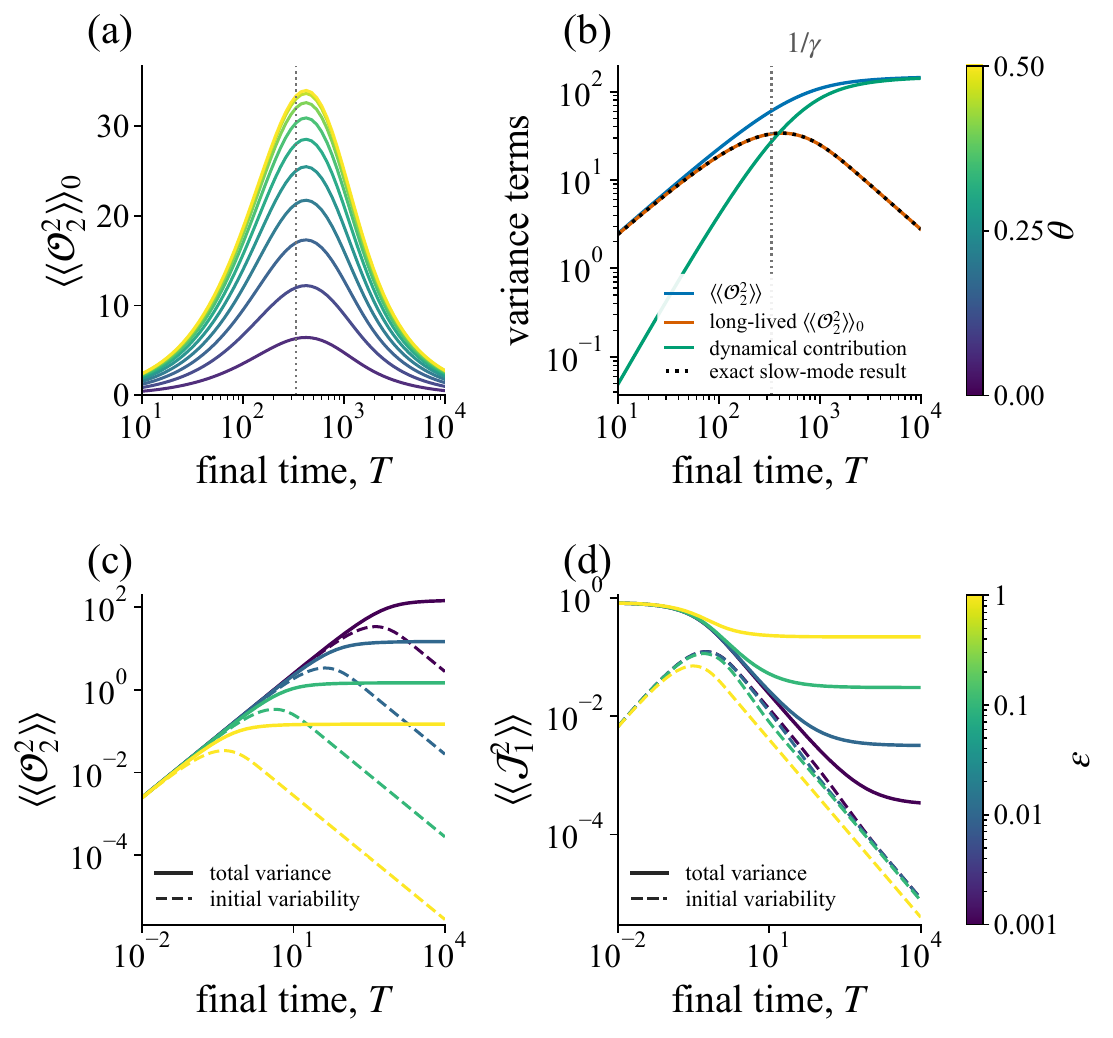}
    \caption{Observable-dependent initial variability in the three-state molecular motor at $\Delta\mu=1$. Panels (a) and (b) use $\varepsilon=10^{-3}$ and hence $\gamma=3\times10^{-3}$. (a) Initial variability $\llangle\mathcal O_2^2\rrangle_0$ of the occupation time $\mathcal O_2=\int_0^Tdt\,\delta_{n_t,2}$ for preparations $\bm p(0)=(\theta,1-\theta,0)^\intercal$, with $0\leq\theta\leq1/2$; the gray dotted line marks $T=1/\gamma$. (b) Total variance, initial variability, and the dynamically generated contribution in \cref{eq:var-FRR} for $\theta=1/2$. The black dotted curve is the exact result $(1-e^{-\gamma T})^2/(4T\gamma^2)$, and the gray dotted line again marks $T=1/\gamma$. (c),(d) Total variance (solid) and initial variability (dashed) for $\theta=1/2$ and $\varepsilon\in\{10^{-3},10^{-2},10^{-1},1\}$, with color indicating $\varepsilon$: (c) $\mathcal O_2$ and (d) the signed current $\mathcal J_1=\int_0^T(dk_1-dk_{-1})$.}
    \label{fig:three-states}
\end{figure}

\blue{To illustrate this regime we consider a three-state molecular motor comprising the reactions $\text{F}+\text{X}_1\rightleftharpoons\text{W}+\text{X}_3$, $\text{X}_1\rightleftharpoons\text{X}_2$, and $\text{X}_2\rightleftharpoons\text{X}_3$, labeled $1$, $2$, and $3$, respectively. The chemical-potential difference $\Delta\mu=\mu_F-\mu_W$ of the chemostatted species $\text{F}$ and $\text{W}$ drives the cycle out of equilibrium. Reaction $1$ is fast, with $W_1=e^{\Delta\mu/2}$ and $W_{-1}=e^{-\Delta\mu/2}$, while the other two reactions are slow, with $W_{\pm2}=W_{\pm3}=\varepsilon\ll1$. In this regime, the spectral gap is $\gamma=3\varepsilon$.}
\blue{For the occupation time $\mathcal O_2=\int_0^Tdt\,\delta_{n_t,2}$, \cref{fig:three-states}(a) shows the initial variability for $\bm p(0)=(\theta,1-\theta,0)^\intercal$, with $0\leq\theta\leq1/2$. It is largest for the equal mixture $\theta=1/2$. The fast mode has zero overlap with this observable, so \cref{eq:variability-relax} is exact with $A_{O_2}=\theta(1-\theta)$ and $\alpha=0$; see \cref{subsec:three-state-relaxation-details}. For $\theta=1/2$, \cref{fig:three-states}(b) compares the initial variability with the total variance and the dynamically generated contribution in \cref{eq:var-FRR}. At $T\sim1/\gamma$, the initial variability is of order $1/\gamma$ and dominates the dynamically generated contribution up to this timescale. As $\varepsilon$ increases, the long-lived enhancement of $\mathcal O_2$ weakens, as shown in \cref{fig:three-states}(c).}
\blue{The effect depends on the observable. For the current through the open reaction,
$\mathcal J_1=\int_0^T(dk_1-dk_{-1})$, the slow-mode prefactor scales as
$A_{J_1}\sim\gamma^2$, while the fast relaxation rate and its overlap with the
current remain $\mathcal O(1)$. Consequently, as $\gamma\to0$ at fixed $T$, the
full initial variability remains $\mathcal O(1)$. In the joint limit
$\gamma T=\mathcal O(1)$, both modal contributions to the excess conditional
current remain $\mathcal O(1)$, so the $1/T$ normalization in
\cref{eq:Omega} gives
$\llangle\mathcal J_1^2\rrangle_0=\mathcal O(\gamma)=\mathcal O(T^{-1})$.
Thus, unlike $\mathcal O_2$, the current exhibits no $1/\gamma$ enhancement;
see \cref{fig:three-states}.
}

\blue{
\textit{Frequency domain---}}We now consider the autonomous dynamics initialized in the
steady state $\bm p(0)=\bm\pi$. The dynamical FRR then reads
\begin{align}
\label{eq:auto-FRR-window-main}
    \langle\!\langle \mathcal Q^2\rangle\!\rangle
    &=
    \langle\!\langle \mathcal Q^2\rangle\!\rangle_0
    +\sum_e
    \frac{1}{j_e^{\mathrm{ss}}
    [\partial_{\lambda_e}\ln W_e]^2}
    \frac{1}{T}\int_0^T d\tau\,
    R_{\lambda_e}^2(\tau,T)\,.
\end{align}
\blue{Since fluctuation--response relations are often formulated in
the frequency domain, we present two distinct frequency 
formulations of the dynamical FRR.}

First, for a fixed observation time $T$, we expand each response kernel
in a Fourier series with discrete frequencies $\omega_k=2\pi k/T$,
defining
$
\hat R_{\lambda_e}^{\,k}(T)
\equiv
\frac{1}{T}\int_0^T d\tau\,
R_{\lambda_e}(\tau,T)e^{i\omega_k\tau}.
$
Applying Parseval's identity to \cref{eq:auto-FRR-window-main} yields the
exact finite-window spectral FRR (see
\cref{subsec:finite-window-response-spectrum}):
\begin{align}
\label{eq:FRR-Fourier-window-main}
    \langle\!\langle \mathcal Q^2\rangle\!\rangle
    &=
    \langle\!\langle \mathcal Q^2\rangle\!\rangle_0
    +\sum_e
    \frac{1}{j_e^{\mathrm{ss}}
    [\partial_{\lambda_e}\ln W_e]^2}
    \sum_{k=-\infty}^{\infty}
    \left|\hat R_{\lambda_e}^{\,k}(T)\right|^2\,.
\end{align}
Keeping only the zero-frequency contribution, $k=0$, in
\cref{eq:FRR-Fourier-window-main} yields the tighter FRI in \cref{eq:FRI-zero-mode}~\cite{kwon2024fluctuation}.

\blue{Second, we consider a harmonically weighted observable as the finite-window Fourier transform 
$\mathcal{Q}^\omega_T = \int_0^T\,e^{-i\omega t}d\mathcal{Q}$
with an arbitrary
real frequency $\omega$ and subsequently take the stationary long-time
limit, $T\to\infty$. This formulation yields a continuous-frequency
power spectrum and the associated susceptibility, while the
contribution from the initial variability vanishes; see
\cref{subsec:stationary-power-spectrum}. We thereby recover the
power-spectrum FRR of Ref.~\cite{kwon2026nonequilibrium}, the FRI of
Ref.~\cite{zheng2026thermodynamic}, and the frequency-domain response
property of Ref.~\cite{voits2026graph}. The two formulations of the finite-frequency domain are
complementary consequences of the same dynamical FRR~\cref{eq:auto-FRR-window-main}.
}

\textit{Periodic driving---}\blue{Now we discuss an important case of nonautonomous driving simplifying \cref{eq:var-FRR}, namely a periodic driving with period $T_p$ and an observable accumulated over the interval $[0,T_M]$, where $T_M=MT_p$ and $M$ is large. After some transient, the system will settle in Floquet state, $\bm p^{\mathrm F}(\tau+T_p)=\bm p^{\mathrm F}(\tau)$~\cite{barato2018current}, where response, $R_{\lambda_e}^{\mathrm F}(\tau)$, and current, $j_e^{\mathrm F}(\tau)=W_e(\tau)p_{s(e)}^{\mathrm F}(\tau)$, become $T_p$-periodic. 
Defining $\llangle\mathcal Q^2\rrangle_{\mathrm F}
\equiv\lim_{M\to\infty} \llangle\mathcal Q^2[\omega_{0:T_M}]\rrangle$, and since transients and initial variability contribute only $\mathcal O(M^{-1})$, \cref{eq:var-FRR} becomes
\begin{align}
\label{eq:var-Floquet}
\llangle\mathcal Q^2\rrangle_{\mathrm F}
=\frac{1}{T_p}\int_0^{T_p}d\tau\sum_e
\frac{[R_{\lambda_e}^{\mathrm F}(\tau)]^2}
{j_e^{\mathrm F}(\tau)[\partial_{\lambda_e}\ln W_e(\tau)]^2}\,,
\end{align}
The derivation and examples based on a driven three-state model and the no-pumping theorem~\cite{rahav2008directed} are detailed in \cref{sec:periodic}.
}

\textit{Specific local perturbations---}\blue{We now derive FRRs underpinning the hierarchy shown in \cref{fig:hierarchy} by considering} 
the parameterization of the transition rates,
\begin{align}
\label{eq:rates-model}
    W_{e}(t) = \exp\big[B_e(t; \varepsilon) + S_e(t; \eta)/2 + E_{s(e)}(t; \theta)\big]\,,
\end{align}
where $B_e \blue{= B_{-e}}$ and $S_e = \blue{- S_{-e}}$ are the symmetric and antisymmetric edge parameters, respectively, and $E_{n}$ are the vertex parameters. 
In stochastic thermodynamics, $\ln W_{e}/W_{- e}=S_e - \bm{\Delta}_e \cdot \boldsymbol{E}$ is the entropy change in the reservoir resulting from the transition $e$, where $S_e$ are the nonconservative contributions and $E_n$ are suitable energy potentials \cite{rao2018conservation, falasco2023macroscopic}.
In turn, $B_e$ can be seen as kinetic barriers along $e$ (controlled by enzymes in metabolic reactions~\cite{Wachtel_2018} or gate voltages in nanoelectronics~\cite{freitas2021stochastic}).
For later purposes, we also introduce specific types of parameters, $\{\varepsilon, \eta, \theta \}$, that act on several edges and vertices. 

For energy perturbation, set $\lambda(\tau)=E_n(\tau)$ in \cref{eq:kernel-general}.
\Cref{eq:rates-model} gives
$\partial_{E_n}\ln W_e=\delta_{n,s(e)}$ because $E_n$ affects precisely the rates whose oriented edges leave state $n$.
Consequently,
\begin{align}
\label{eq:response-E}
    R_{E_n}(\tau, T) = \sum_e \blue{\delta_{ns(e)} j_e(\tau)}\varphi_e(\tau,T)\,,
\end{align}
which is a dynamical generalization of the static energy responses~\cite{owen2020universal,aslyamov2024nonequilibrium}. 
For perturbations in $\lambda(\tau) \in \{B_e(\tau), S_e(\tau)\}$, using \cref{eq:rates-model} to calculate $\partial_\lambda \ln W_e(t)$ in \cref{eq:kernel}, we find
\begin{align}
\label{eq:response-BS-general}
    R_{B_e}(\tau, T) &= \varphi_e(\tau, T) j_e(\tau) + \varphi_{-e}(\tau, T) j_{-e}(\tau)\,,\\
    R_{S_e}(\tau, T) &= \frac{1}{2}[\varphi_e(\tau, T) j_e(\tau) -  \varphi_{-e}(\tau, T) j_{-e}(\tau)]\,,
\end{align}
\blue{which hold for arbitrary edge weights $x_e(t)$.}

\blue{
For the remainder of the Letter, we assume antisymmetric weights $x_e(t) = -x_{-e}(t)$ implying $\varphi_e(t) = -\varphi_{-e}(t)$. This corresponds to considering state and current-based observables.}

\blue{First, under this assumption, substituting $\lambda_{e}(t) = W_{e}(t)$ into \cref{eq:kernel}, we find}
\begin{align}
\label{eq:mutual-linearity}
    \frac{R_{W_{e}}(\tau,T)}{R_{W_{-e}}(\tau,T)} = -\frac{p_{s(e)}(\tau)}{p_{t(e)}(\tau)}\,.
\end{align}
In the steady-state limit, this result reduces to the mutual-linearity corollary of Ref.~\cite{bebon2026mutual} 
\blue{and, as shown in 
\cref{subsec:stationary-power-spectrum}, to the frequency-domain response relation~\cite{voits2026graph}.}

\blue{Second, \cref{eq:response-BS-general} simplifies to $R_{B_e}(\tau, T) = \varphi_e(\tau, T) I_e(\tau)$ and $R_{S_e}(\tau, T) = \frac{1}{2}\varphi_e(\tau, T) a_e(\tau)$.} 
These response kernels satisfy
\begin{align}
\label{eq:ratio}
    \frac{R_{B_e}(\tau, T)}{R_{S_e}(\tau, T)} = \frac{2 I_e(\tau)}{a_e(\tau)}\,,
\end{align}
extending similar relations~\cite{aslyamov2024nonequilibrium,kwon2024fluctuation} to nonautonomous dynamics.  

\blue{Third, by expressing $\varphi_e$ in terms of $S$-response kernels and inserting the result into \cref{eq:FRR},}
we obtain the dynamical generalization of the static FRRs~\cite{aslyamov2025frr, ptaszynski2024frr, ptaszynski2025frr-mixed}:
\begin{align}
\label{eq:FRR-BS}
      \langle\!\langle \mathcal{Q}, \mathcal{Q}' \rangle\!\rangle =\langle\!\langle\mathcal{Q}, \mathcal{Q}' \rangle\!\rangle_{0}+  &\int_0^T \frac{dt}{T} \sum_{e>0} \frac{4R_{S_e}(t, T)R'_{S_e}(t, T)}{a_e(t)}
    \,,
\end{align}
as well as the barriers-perturbations counterpart using the substitution $R_{S_e}=a_e R_{B_e}/(2I_e)$ [\cref{eq:ratio}]. 
We note that there is no direct vertex-parameter analog of \cref{eq:FRR-BS} for perturbations in $E_n$. 

\blue{
\textit{Thermodynamic inequalities---}Thermodynamic inequalities relate measurable fluctuations and responses to less accessible quantities such as entropy production and dynamical activity (traffic). However, the landscape of existing bounds is difficult to navigate because they have been derived using a variety of methods. The DFRR \eqref{eq:FRR-BS} with \cref{eq:ratio} unifies many of these results and yields tighter forms that account for variability in the initial state; see \cref{fig:hierarchy}. Here, we derive dynamical R-TURs and R-KURs. In the End Matter, we further show how the same framework sharpens several known bounds.
}

We start with a local version of the dynamical R-TUR and R-KUR defined in terms of edge perturbations $\{B_e, S_e\}$. Using \cref{eq:FRR-BS} for $\mathcal{Q}=\mathcal{Q}'$ and applying the Sedrakyan integral inequality, we find:
\begin{align}
\label{eq:edge-R-TUR}
    \overline{\sigma}^\text{ps}_e(T) &\geq \frac{\overline{R}^2_{B_e}(T)}{\langle\!\langle \mathcal{Q}^2 \rangle\!\rangle - \langle\!\langle \mathcal{Q}^2 \rangle\!\rangle_{0}}\,,\,
    \overline{a}_e(T) \geq \frac{4 \overline{R}^2_{S_e}(T)}{\langle\!\langle \mathcal{Q}^2 \rangle\!\rangle - \langle\!\langle \mathcal{Q}^2 \rangle\!\rangle_{0}}\,.
\end{align}
\blue{These bounds have remarkably simple forms, but require the microscopic responses $\overline{R}_{B_e}$ and $\overline{R}_{S_e}$.} 

Next, we derive the dynamical R-TUR for perturbation of the kinetic parameter $\varepsilon$ that acts on several edges with the response 
$d_\varepsilon Q(0, T) = \int_0^T d\tau \sum_{e>0} R_{B_e}(\tau, T) \partial_\varepsilon B_e(\tau)$,
where the response kernel is calculated along the unperturbed driven evolution. 
Using the FRR~\cref{eq:FRR-BS} and algebraic inequalities (see \cref{eq:R-TUR-proof} in the End Matter), we find the dynamical generalization of the static R-TUR~\cite{ptaszynski2024dissipation,aslyamov2025frr}:
\begin{align}
\label{eq:R-TUR}
    \overline{\sigma}(T) 
    \geq \frac{2}{T^2 b^2_\text{max}}\frac{\big[ d_\varepsilon Q(0,T) \big]^2}{\langle\!\langle \mathcal{Q}^2 \rangle\!\rangle-\langle\!\langle \mathcal{Q}^2 \rangle\!\rangle_{0}} 
    \geq \frac{2\big[ d_\varepsilon Q(0,T) \big]^2}{T^2 b^2_\text{max}\langle\!\langle \mathcal{Q}^2 \rangle\!\rangle}\,,
\end{align}
where $\overline{\sigma}(T)= \sum_{e>0} \overline{\sigma}_e(T)$ is the time-averaged entropy production and $b_\text{max} = \max_{e,t}|\partial_\varepsilon B_e(t)|$ is the model-dependent parameter. 
\blue{In the End Matter, we show that \cref{eq:R-TUR} for a special case of kinetic perturbations $B^\varepsilon_e = B_e + \ln(1+ \varepsilon)$ with $b_\text{max} = 1$ results in the tighter form of time-dependent driving TUR \cite{koyuk2020thermodynamic}.}

To derive dynamical R-KUR, we introduce the physically motivated decomposition of the nonconservative term as $S_e(t;\eta) = \sum_{\gamma}X_{e\gamma}(t)F_\gamma(t;\eta)$, where $F_\gamma$ denotes a set of thermodynamic forces, while $X_{e\gamma}$ specifies how force $\gamma$ contributes to the nonconservative term on transition $e$ \cite{rao2018detailed}. 
Two standard choices are particularly useful. In cycle/co-cycle representation~\cite{rao2018detailed}, $F_\gamma$ are the affinities of a chosen set of independent cycles, and $X_{e\gamma}$ is the edge--cycle incidence matrix, encoding whether edge $e$ belongs to cycle $\gamma$ and with which orientation. In the thermodynamic-force representation, $F_\gamma$ are the fundamental forces imposed by the reservoirs, and $X_{e\gamma}$ maps those reservoir forces onto the transitions they drive~\cite{rao2018conservation}. 
When all forces are set to zero, $F_\gamma=0$ for all $\gamma$, the dynamics is said to be detailed balanced. 
We now derive dynamical inequalities for perturbations in $\eta$. Using \cref{eq:FRR-BS} and Sedrakyan's inequality [\cref{eq:R-KUR-proof}], we prove the dynamical R-KUR:
\begin{align}
\label{eq:R-KUR}
    &\langle\!\langle \mathcal{Q}^2 \rangle\!\rangle -\langle\!\langle \mathcal{Q}^2 \rangle\!\rangle_{0} \geq  \frac{4\Big[\int_0^T \tfrac{d\tau}{T} \sum_{\gamma} R_{F_\gamma}(\tau, T)\partial_\eta F_\gamma(\tau)\Big]^2}{\int_0^T \tfrac{d\tau}{T} \sum_{\gamma,\gamma'}A_{\gamma\gamma'}(\tau)
\partial_\eta F_\gamma(\tau)\partial_\eta F_{\gamma'}(\tau)}\,,
\end{align}
where $R_{F_\gamma}(\tau,T)\equiv \delta Q(0,T)/\delta F_\gamma(\tau)$ are the force responses and where $A_{\gamma\gamma'}(\tau) = \sum_{e>0} a_e(\tau) X_{e\gamma}(\tau)X_{e\gamma'}(\tau)$ is the weighted activity matrix. In the End Matter, we show two special cases of dynamical R-KUR \eqref{eq:R-KUR}: first, we find the optimal R-KUR maximizing the right-hand side of \cref{eq:R-KUR}; second, by retaining initial variability in \cref{eq:R-KUR} we derive a tighter form of the linear FRI from~\cite{dechant2020fluctuation}.

\blue{\textit{Connection to FDTs---}The dynamical FRR extends the standard FDT to arbitrary time-dependent and nonconservative driving. For detailed-balance dynamics with a time-independent equilibrium preparation, \cref{eq:FRR} reduces to a symmetrized response relation. In the autonomous case, it further yields the standard FDT and Onsager reciprocity; see \cref{sec:KuboOnsager}. 
For a general dynamics, the FRRs also recover the nonequilibrium FDTs based on the covariance representation of responses~\cite{agarwal1972fluctuation,seifert2010fluctuation,zheng2025nonlinear}; see \cref{sec:discrepancy}.
}

\blue{\textit{Conclusions---}The dynamical FRRs derived in this Letter provide exact relations between fluctuations and response that apply to any observable of any autonomous or non-autonomous Markov jump process, for arbitrary initial preparations and observation times. Its central structural feature is the decomposition of finite-time covariances into a contribution inherited from the initial preparation and a contribution generated by the subsequent dynamics, expressed in terms of response kernels. This structure provides a unified understanding of numerous identities and bounds that have recently been derived using a variety of methods and sharpens their finite-time forms by retaining the initial-variability contribution, as summarized in \cref{fig:hierarchy}.}

\blue{A natural next step is to extend dynamical FRRs beyond finite-state jump processes, both to diffusive dynamics and to macroscopic limits. A related decomposition of the current variance into initial and dynamically generated contributions has recently appeared in macroscopic fluctuation theory for diffusive relaxation~\cite{suzuki2026non}. Such extensions would enable the application of dynamical FRRs to two distinct settings in which slow relaxation and preparation dependence play important roles. 
The first is nonequilibrium phase transitions, where the closing of a relaxation gap leads to increasingly slow dynamics and may render the contribution from initial variability long-lived. Indeed, Ref.~\cite{tamano2026universal} used a macroscopic counterpart of our Floquet variance relation~\eqref{eq:var-Floquet} to characterize the critical divergence of entropy-production fluctuations near a Hopf bifurcation in chemical reaction networks. 
The second is aging and glassy relaxation in disordered systems, often even under detailed-balance dynamics, where slow modes retain memory of the preparation and produce long-time violations of equilibrium fluctuation--dissipation relations~\cite{cugliandolo1993analytical}. In both settings, dynamical FRRs could help disentangle the contributions of preparation-induced variability and dynamically generated fluctuations to observables.}

\begin{acknowledgments}
T.A. and M.E. are funded by the Fonds National de la Recherche-FNR, Luxembourg: project ThermoElectroChem (C23/MS/18060819) and NEQPHASETRANS (C24/MS/18933049), respectively.
The authors thank Krzysztof Ptaszy\'{n}ski for helpful comments on the frequency-domain formulations. 
\end{acknowledgments}

\newpage

\begin{center}
  \large \bf End Matter
\end{center}


\textit{Derivation of \cref{eq:R-TUR,eq:R-KUR}---}Here we derive \cref{eq:R-TUR,eq:R-KUR}. We start with the R-TUR. Using \cref{eq:FRR-BS} together with $ R_{S_e}=a_eR_{B_e}/(2I_e)$, we find
\begin{align}
\label{eq:R-TUR-proof}
    &\langle\!\langle \mathcal{Q}^2 \rangle\!\rangle -\langle\!\langle \mathcal{Q}^2 \rangle\!\rangle_{0} 
    = \frac{1}{T}\int_0^Tdt\sum_{e>0}
\frac{\left[b_e(t)I_e(t)\varphi_e(t,T)\right]^2}{b_e^2(t)\sigma^\text{ps}_e(t)} \nonumber\\
&\geq  \frac{1}{T}\int_0^T  dt \frac{\Big(\sum_{e>0} b_e(t) I_e(t)\varphi_e(t, T)\Big)^2}{\sum_{e>0} b^2_e(t) \sigma^\text{ps}_e(t)} \nonumber\\
&\geq \frac{1}{T^2}\frac{\big(d_\varepsilon Q\big)^2}{b^2_\text{max}\overline \sigma^\text{ps}(T)} \geq \frac{2}{T^2}\frac{\big(d_\varepsilon Q\big)^2}{b^2_\text{max}\overline \sigma(T)}\,,
\end{align}
where $\sigma_e^\text{ps}(t) = I^2_e(t)/a_e(t)$ and where $b_e(t) = \partial_{\varepsilon}B_e(t)$ with $b_\text{max} = \max_{e,t}|b_e(t)|$. 
For the first inequality, we used Sedrakyan’s series inequality $\sum_{n}f^2_n/g_n\geq (\sum_n f_n)^2/\sum_n g_n$ for  all $f_n$ and positive $g_n>0$; for the second inequality, we used Sedrakyan’s integral inequality with $\int_0^T dt\sum_{e>0} b^2_e(t)\sigma^\text{ps}_e(t) \leq b^2_\text{max} T \overline \sigma^\text{ps}(T)$, for the third inequality, we used  $\sigma^\text{ps}_e \leq \sigma_e/2$. We now derive the R-KUR:
\begin{align}
\label{eq:R-KUR-proof}
    &\langle\!\langle \mathcal{Q}^2 \rangle\!\rangle -\langle\!\langle \mathcal{Q}^2 \rangle\!\rangle_{0} = \frac{1}{T}\int_0^T dt \sum_{e>0} \frac{(\partial_\eta S_e(t) a_e(t) \varphi_e(t, T))^2}{[\partial_\eta S_e(t)]^2 a_e(t)} \nonumber\\
    &\geq  \frac{4}{T}\frac{\Big[\int_0^T d\tau \sum_{\gamma} R_{F_\gamma}(\tau, T)\partial_{\eta} F_\gamma(\tau)\Big]^2}{\int_0^Td\tau\sum_{\gamma,\gamma'}A_{\gamma\gamma'}(\tau)
\partial_{\eta} F_\gamma(\tau)\partial_{\eta} F_{\gamma'}(\tau)}\,,
\end{align}
where we used Sedrakyan’s inequalities and $\partial_\eta S_e = \sum_{\gamma} X_{e \gamma} \partial_\eta F_\gamma$.

\textit{Tighter form of the time-dependent driving TUR---}As a special case, we consider a ``virtual'' perturbation of all barriers, $B^\varepsilon_e = B_e + \ln(1+ \varepsilon)$, implying $b_\text{max} = 1$ and a shift $W^\varepsilon_{\pm e} = (1+\varepsilon)W_{\pm e}$ as in Refs.~\cite{di2018kinetic, koyuk2020thermodynamic}.
Introducing the speed of driving protocol $v$, we notice that the virtually perturbed system is equivalent to the unperturbed dynamics observed at rescaled time $t^{\varepsilon} = (1 + \varepsilon)t$ under the protocol with the rescaled speed $v^{\varepsilon}=v/(1+\varepsilon)$ such that  $W (v^\varepsilon t^\varepsilon) = W(v t)$. Then using $O^\varepsilon(0,T; v) = (1+\varepsilon)^{-1}O(0, T^{\varepsilon};v^{\varepsilon})$ and $J^\varepsilon(0,T; v) = J(0, T^{\varepsilon};v^{\varepsilon})$, we find the response
\begin{align}
\label{eq:d_h}
     d_\varepsilon Q^{\varepsilon}(0,T)\big|_{\varepsilon = 0} = -O(0, T) + (T \partial_T - v \partial_v) Q(0,T)\,.
\end{align}
Here the observable weights are assumed to be constant.

Using \cref{eq:R-TUR} for current observables, $\mathcal{Q}=\mathcal{J}$, and kinetic response \eqref{eq:d_h} with $b_\text{max} = 1$, we find 
\begin{align}
\label{eq:KS-TUR}
    \overline \sigma(T) 
    \geq \frac{2}{T^2}\frac{\big[(T \partial_T - v \partial_v)J(0,T)\big]^2}{\langle\!\langle \mathcal{J}^2 \rangle\!\rangle-\langle\!\langle \mathcal{J}^2 \rangle\!\rangle_{0}}
    \geq \frac{2\big[(T \partial_T - v \partial_v)J(0,T)\big]^2}{T^2\langle\!\langle \mathcal{J}^2 \rangle\!\rangle}\;,
\end{align} 
where the last looser inequality is the time-dependent driving TUR \cite{koyuk2020thermodynamic}, which, for autonomous relaxation ($\partial_v J(0,T)=0$), further reduces to $\overline \sigma(T)\geq 2[\bm{x}(T)\cdot\bm{I}(T)]^2/\langle\!\langle \mathcal{J}^2 \rangle\!\rangle$ known from Ref.~\cite{liu2020thermodynamic}. We see that the presence of the initial variability term makes our bounds always tighter.

To illustrate the gain from retaining the initial variability, we return to the autonomous three-state model and monitor the current $\mathcal J_1$, starting from $\bm p(0)=(\theta,1-\theta,0)^\intercal$. In this case, the two inequalities in~\eqref{eq:KS-TUR} give
$\mathcal B_{\rm tight}=2I_1(T)^2/[\llangle\mathcal J_1^2\rrangle-\llangle\mathcal J_1^2\rrangle_0]$ and $\mathcal B_{\rm TUR}=2I_1(T)^2/\llangle\mathcal J_1^2\rrangle$, where $I_1(T)$ is the instantaneous mean current at the final time. Their ratio has the transparent form
$\mathcal B_{\rm tight}/\mathcal B_{\rm TUR}=1/[1-\llangle\mathcal J_1^2\rrangle_0/\llangle\mathcal J_1^2\rrangle]$: the improvement is therefore controlled directly by the fraction of the current variance inherited from the initial ensemble.

\begin{figure}
    \centering
    \includegraphics[width=\linewidth]{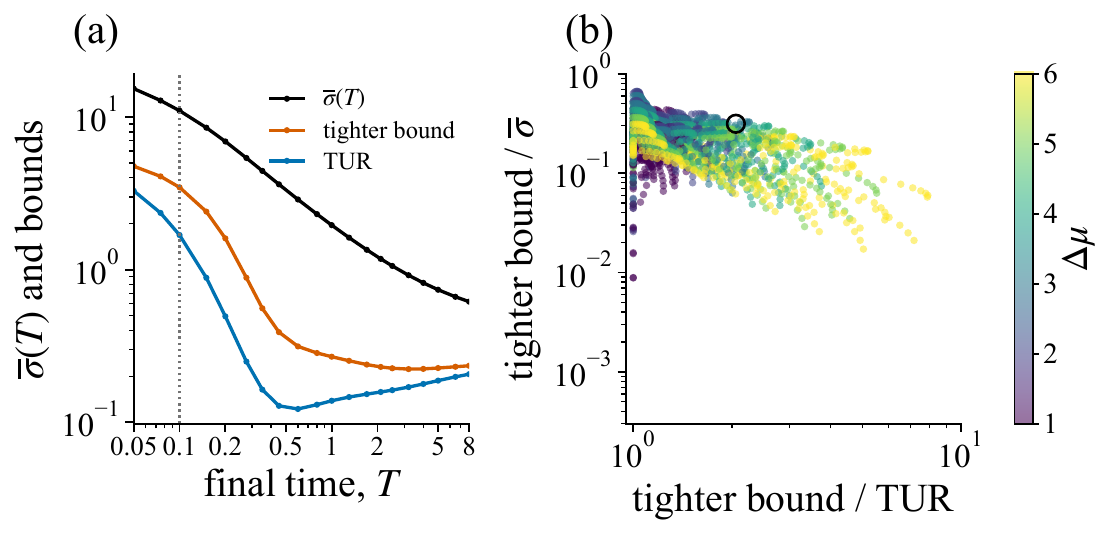}
    \caption{Finite-time EPR inference in the autonomous three-state model. (a) Time-averaged EPR $\overline\sigma(T)$, initial-variability-aware bound $\mathcal B_{\rm tight}(T)=2I_1(T)^2/[\llangle\mathcal J_1^2\rrangle-\llangle\mathcal J_1^2\rrangle_0]$, and conventional TUR bound $\mathcal B_{\rm TUR}(T)=2I_1(T)^2/\llangle\mathcal J_1^2\rrangle$ for the reference parameters. The dotted line marks $T_\ast=0.1$. (b) Parameter sweep shown in terms of the improvement factor $\mathcal B_{\rm tight}/\mathcal B_{\rm TUR}$ and the inferred EPR fraction $\mathcal B_{\rm tight}/\overline\sigma$. Color denotes the affinity $\Delta\mu$; the open circle marks the reference point from panel~(a). The complete parameter grid is given in Ref.~\cite{parameters-1}.}
    \label{fig:epr-bounds-sweep}
\end{figure}

For the reference parameters in \cref{fig:epr-bounds-sweep}(a), the tighter bound recovers $31.4\%$ of $\overline\sigma(T)$ at $T_\ast=0.1$, compared with $15.3\%$ for the conventional TUR, an improvement by a factor of about $2.06$. To test whether this gain requires fine tuning, \cref{fig:epr-bounds-sweep}(b) collects $2376$ parameter combinations spanning three kinetic and energetic backgrounds, six affinities, eleven initial mixtures, and twelve final times. Moving to the right means that initial variability produces a larger improvement over the TUR, whereas moving upward means that the tighter bound captures a larger fraction of the actual EPR. The sweep contains broad regimes in which the tighter bound substantially improves upon the conventional TUR while still recovering an appreciable fraction of the EPR.

\textit{Optimal TUR---}We notice that \cref{eq:d_h} in the limit $T \to \infty$ simplifies to $T^{-1} d_\varepsilon Q = \bm x \cdot \bm I^\text{ss} = \sum_{e>0} x_e I^\text{ss}_e$. 
Inserting this result into \cref{eq:R-TUR} and introducing the steady-state EPR $\sigma_\text{ss}\equiv \overline\sigma(\infty)$, we obtain 
\begin{align}
\label{eq:EPR-ss-bound}
   \sigma_\text{ss} \geq \frac{2 [\bm{x}\cdot\bm{I}_{\text{ss}}]^2}{\langle\!\langle \mathcal{Q}^2 \rangle\!\rangle_\text{ss}}\,,
\end{align} 
which is known as an optimal TUR-type inequality~\cite{shiraishi2021optimal}.
For $\mathcal{Q} = \mathcal{J} - \alpha \mathcal{O}$, the homogeneous kinetic response in the steady-state limit is independent of $\alpha$, while the variance can be minimized over $\alpha$.
We find 
$\min_\alpha\langle\!\langle \mathcal{Q}^2\rangle\!\rangle_\text{ss} = \langle\!\langle \mathcal{J}^2\rangle\!\rangle_\text{ss}(1-\nu_{\mathcal{J},\mathcal{O}}^2)$, where $\nu_{\mathcal{J},\mathcal{O}}^2 = \langle\!\langle \mathcal{J}, \mathcal{O}\rangle\!\rangle^2_\text{ss}/(\langle\!\langle \mathcal{J}^2\rangle\!\rangle_\text{ss} \langle\!\langle \mathcal{O}^2\rangle\!\rangle_\text{ss})$. Inserting this result in \cref{eq:EPR-ss-bound}, we find the main result of \cite{dechant2021improving}.

\textit{Special cases of dynamical R-KUR~\eqref{eq:R-KUR}---}A first special case of \cref{eq:R-KUR} is the optimal R-KUR.
The bound \eqref{eq:R-KUR} holds for an arbitrary $\partial_\eta F_\gamma(t)$. We find $\partial_\eta \bm F^*(t)$ that maximizes the right-hand side of \cref{eq:R-KUR}. Solving quadratic optimization~\cite{vu2020entropy}, we find $\partial_\eta \bm F^*(t) \propto A^{+} \bm R_{F}(t, T)$, where $\bm R_{F} = (\dots,R_{F_\gamma},\dots)^\intercal$ and where $A^{+}$ denotes the Moore--Penrose inverse; when $A$ is positive definite $A^+ = A^{-1}$. We insert $\partial_\eta \bm F^*(t)$ in \cref{eq:R-KUR}:
\begin{align}
    \label{eq:R-KUR-opt}
    \langle\!\langle \mathcal{Q}^2 \rangle\!\rangle \geq& \langle\!\langle \mathcal{Q}^2\rangle\!\rangle_{0}\\ \nonumber
    &+ \frac{4}{T}\int_0^T d\tau\sum_{\gamma,\gamma'} R_{F_\gamma}(\tau,T) [A^{+}(\tau)]_{\gamma\gamma'}R_{F_{\gamma'}}(\tau,T)\,.
\end{align}
We see that moving from edge perturbations to force perturbations, the equality~\eqref{eq:FRR-BS} has become the inequality \eqref{eq:R-KUR-opt}. 

A second special case of \cref{eq:R-KUR} is obtained by combining \cref{eq:R-KUR} with the result that the Kullback–Leibler divergence between perturbed and unperturbed probability densities can be expressed as $D_\text{KL}(\mathbb{P}^\text{pert}\Vert\mathbb{P}) = D^{(2)}_\text{KL}(\mathbb{P}^\text{pert}\Vert\mathbb{P})+\mathcal O(\delta F^3)$ with $D^{(2)}_\text{KL}(\mathbb{P}^\text{pert}\Vert\mathbb{P}) = \tfrac{1}{8}\int d\tau \sum_{\gamma,\gamma'} A_{\gamma\gamma'}(\tau)\delta F_\gamma(\tau)\delta F_{\gamma'}(\tau)$~\cite{dechant2020fluctuation}, so that
\begin{align}
\label{eq:Dechant-Sasa}
    D^{(2)}_\text{KL}(\mathbb{P}^\text{pert}\Vert\mathbb{P}) \geq \frac{1}{2T}\frac{\big(\delta Q \big)^2}{\langle\!\langle \mathcal{Q}^2 \rangle\!\rangle-\langle\!\langle \mathcal{Q}^2 \rangle\!\rangle_{0}}\,,
\end{align}
where $\delta Q\equiv \sum_{\gamma}\int_0^T d\tau R_{F_\gamma}(\tau,T)\delta F_\gamma(\tau) + \mathcal{O}(\delta \bm F^2)$. Ignoring the initial variability, this result reduces to the looser linear FRI from~\cite{dechant2020fluctuation}.

\newpage
\onecolumngrid
\appendix
\crefalias{section}{appendix}
\section{Derivation of dynamical FRRs}
\label{sec:DFRR}

We first derive Eq.~(9a) from the main text which is 
\begin{align}
\label{eq:general-kernel-SM}
    \frac{\delta Q(\tau,T)}{\delta\lambda(\tau)}
    &=  \sum_e \varphi_e(\tau,T)j_e(\tau)
    \frac{\partial\ln W_e(\tau)}{\partial\lambda(\tau)}\,.
\end{align}
Recall that $Q_n(\tau,T)$ is the conditional mean of the observable accumulated over $[\tau,T]$, given that the system is in state $n$ at time $\tau$.
Consider a sufficiently short interval $[\tau,\tau+\Delta t]$, during which either no jump or one jump along an oriented edge $e$ occurs.
The corresponding probabilities are $1+\Delta t\,W_{nn}(\tau)$ and $\delta_{n,s(e)}W_e(\tau)\Delta t$, respectively, where
$W_{nn}(\tau)=-\sum_e\delta_{n,s(e)}W_e(\tau)$.
The conditional mean therefore satisfies
\begin{align}
\label{eq:Q_n-estimate}
    Q_n(\tau,T) ={}& [1+\Delta t\,W_{nn}(\tau)]
    [\Delta t\,o_n(\tau)+Q_n(\tau+\Delta t,T)] +\Delta t\sum_e\delta_{n,s(e)}W_e(\tau)
    [x_e(\tau)+Q_{t(e)}(\tau+\Delta t,T)]
    +\mathcal O(\Delta t^2)\,.
\end{align}

Holding the observable weights fixed, we define the instantaneous functional derivative by
\begin{align}
\label{eq:dQ_ndW_e}
    \frac{\delta Q_n(\tau,T)}{\delta W_e(\tau)}
    &\equiv \lim_{\Delta t\to0}\frac{1}{\Delta t}
    \frac{\partial Q_n(\tau,T)}{\partial W_e(\tau)} = \frac{\partial W_{nn}(\tau)}{\partial W_e(\tau)}Q_n(\tau,T)
    +\delta_{n,s(e)}[x_e(\tau)+Q_{t(e)}(\tau,T)] \nonumber\\
    &= \delta_{n,s(e)}
    [x_e(\tau)+Q_{t(e)}(\tau,T)-Q_{s(e)}(\tau,T)] = \delta_{n,s(e)}\varphi_e(\tau,T)\,.
\end{align}
Here we used $\delta Q_n(\tau+\Delta t,T)/\delta W_e(\tau)=0$ and the definition of $\varphi_e \equiv x_e(\tau)+Q_{t(e)}(\tau,T)-Q_{s(e)}(\tau,T)$.
Since $\bm p(\tau)$ is the unperturbed distribution at the beginning of the interval, averaging \cref{eq:dQ_ndW_e} gives
\begin{align}
\label{eq:dQdW}
    R_{W_e}(\tau,T)
    \equiv \frac{\delta Q(\tau,T)}{\delta W_e(\tau)}
    = \sum_n p_n(\tau)\frac{\delta Q_n(\tau,T)}{\delta W_e(\tau)}
    = p_{s(e)}(\tau)\varphi_e(\tau,T)\,.
\end{align}
For an arbitrary rate parameterization $W(\lambda(\tau))$, the chain rule and
$j_e(\tau)=W_e(\tau)p_{s(e)}(\tau)$ yield \cref{eq:general-kernel-SM} as
\begin{align*}
    \frac{\delta Q(\tau,T)}{\delta\lambda(\tau)}
    &= \sum_e
    \frac{\delta Q(\tau,T)}{\delta W_e(\tau)}
    \frac{\partial W_e(\tau)}{\partial\lambda(\tau)} = \sum_e \varphi_e(\tau,T)j_e(\tau)
    \frac{\partial\ln W_e(\tau)}{\partial\lambda(\tau)}\,.
\end{align*}

Now, we derive the dynamical fluctuation-response relation (FRR).
The scaled covariance follows from the finite-time moment-generating function~\cite{lebowitz1999gallavotti,esposito2007entropy}:
\begin{align}
    \label{eq:scgf}
    \langle\!\langle \mathcal{Q},\mathcal{Q}'\rangle\!\rangle
    &= \frac{1}{T}\left.\partial_{\ell}\partial_{\ell'}
    \ln G(0,T,\ell,\ell')\right|_{\ell=\ell'=0}\,,\nonumber\\
    G(0,T,\ell,\ell')
    &\equiv \sum_{\omega_{0:T}}
    e^{\ell\mathcal{Q}[\omega_{0:T}]+\ell'\mathcal{Q}'[\omega_{0:T}]}
    \mathbb{P}[\omega_{0:T}]\,,
\end{align}
with $G(0,T,0,0)=1$.
We also introduce the conditional generating function and its logarithm,
\begin{align}
    \label{eq:conditional-cgf}
    G_n(t,T,\ell,\ell')
    &\equiv \sum_{\omega_{t:T}}
    e^{\ell\mathcal{Q}[\omega_{t:T}]+\ell'\mathcal{Q}'[\omega_{t:T}]}
    \mathbb{P}[\omega_{t:T}\mid n_t=n]\,,\nonumber\\
    H_n(t,T,\ell,\ell')&\equiv\ln G_n(t,T,\ell,\ell')\,.
\end{align}
Their derivatives at the origin satisfy
\begin{align}
    \label{eq:gf-moments}
    H_n(t,T,0,0)&=0\,,\qquad
    Q_n(t,T)=\partial_\ell H_n(t,T,0,0)\,,\nonumber\\
    Q'_n(t,T)&=\partial_{\ell'}H_n(t,T,0,0)\,,\qquad
    V_n(t,T)=\partial_\ell\partial_{\ell'}H_n(t,T,0,0)\,.
\end{align}
The terminal conditions are $H_n(T,T,\ell,\ell')=0$ and hence
$Q_n(T,T)=Q'_n(T,T)=V_n(T,T)=0$. Using
$G(0,T,\ell,\ell')=\sum_n p_n(0)e^{H_n(0,T,\ell,\ell')}$
in \cref{eq:scgf} gives
\begin{align}
    \label{eq:gf-cov}
    \langle\!\langle \mathcal{Q},\mathcal{Q}'\rangle\!\rangle
    &=\frac{1}{T}\sum_n p_n(0)\Delta Q_n(0,T)\Delta Q'_n(0,T)
    +\frac{v(0,T)}{T}\,,
\end{align}
where $v(t,T)\equiv\sum_n p_n(t)V_n(t,T)$ and
$\Delta Q_n(t,T)\equiv Q_n(t,T)-Q(t,T)$, with the analogous definition for $\Delta Q'_n$.
We denote the first term by
$\langle\!\langle\mathcal Q,\mathcal Q'\rangle\!\rangle_0$; it is the initial variability.
It remains to derive the backward equation for $Q_n(t,T)$ and evaluate $v(0,T)$.

Consider the short interval $[t,t+\Delta t]$.
With the oriented-edge convention, a trajectory either remains in state $n$ or jumps along an edge $e$ with $s(e)=n$.
Thus,
\begin{align}
    \label{eq:gf-short-time}
    G_n(t,T,\ell,\ell')={}&[1+\Delta t\,W_{nn}(t)]
    G_n^{(0)}(t,T,\ell,\ell')+\Delta t\sum_e\delta_{n,s(e)}W_e(t)
    G_n^{(e)}(t,T,\ell,\ell')+\mathcal O(\Delta t^2)\,.
\end{align}
The no-jump and one-jump contributions needed to this order are
\begin{align}
    \label{eq:gf-term-linear}
    G_n^{(0)}(t,T,\ell,\ell')
    ={}&e^{\Delta t[\ell o_n(t)+\ell'o'_n(t)]}
    G_n(t+\Delta t,T,\ell,\ell')\nonumber\\
    ={}&G_n(t,T,\ell,\ell')
    +\Delta t\big\{\partial_tG_n(t,T,\ell,\ell')
    +[\ell o_n(t)+\ell'o'_n(t)]G_n(t,T,\ell,\ell')\big\}
    +\mathcal O(\Delta t^2)\,,\nonumber\\
    G_n^{(e)}(t,T,\ell,\ell')
    ={}&e^{\ell x_e(t)+\ell'x'_e(t)}
    G_{t(e)}(t,T,\ell,\ell')+\mathcal O(\Delta t)\,.
\end{align}
Substituting \cref{eq:gf-term-linear} into \cref{eq:gf-short-time}, using
$W_{nn}(t)=-\sum_e\delta_{n,s(e)}W_e(t)$, and writing $H_n=\ln G_n$, we obtain
\begin{align}
    \label{eq:backward-cgf}
    -\partial_tH_n(t,T,\ell,\ell')={}&\ell o_n(t)+\ell'o'_n(t)\nonumber\\
    &+\sum_e\delta_{n,s(e)}W_e(t)
    \left\{e^{\ell x_e(t)+\ell'x'_e(t)
    +H_{t(e)}(t,T,\ell,\ell')-H_n(t,T,\ell,\ell')}-1\right\}\,.
\end{align}

Taking $\partial_\ell$ of \cref{eq:backward-cgf} at $\ell=\ell'=0$ yields
\begin{align}
    \label{eq:backward-Q}
    -\partial_tQ_n(t,T)
    ={}&o_n(t)+\sum_e\delta_{n,s(e)}W_e(t)
    [x_e(t)+Q_{t(e)}(t,T)-Q_n(t,T)]\nonumber\\
    ={}&[W^\intercal(t)\bm Q(t,T)]_n+\partial_{p_n}\dot Q(t)\,,
\end{align}
where
$\dot Q(t) \equiv \sum_n o_n(t)p_n(t)+\sum_e x_e(t)W_e(t)p_{s(e)}(t)$.
In vector form, \cref{eq:backward-Q} is the backward Kolmogorov equation
\begin{align}
    -\partial_t\bm Q(t,T)
    =W^\intercal(t)\bm Q(t,T)+\bm\nabla_p\dot Q(t)\,,
\end{align}
with $\bm Q(T,T)=0$.

Taking the mixed derivative $\partial_\ell\partial_{\ell'}$ instead gives
\begin{align}
    \label{eq:backward-V}
    -\partial_tV_n(t,T)={}&
    \sum_e\delta_{n,s(e)}W_e(t)
    \varphi_e(t,T)\varphi'_e(t,T)
    +\sum_m W_{mn}(t)V_m(t,T)\,,
\end{align}
where
\begin{align}
    \varphi_e(t,T)&=x_e(t)+Q_{t(e)}(t,T)-Q_{s(e)}(t,T)\,,\nonumber\\
    \varphi'_e(t,T)&=x'_e(t)+Q'_{t(e)}(t,T)-Q'_{s(e)}(t,T)\,.
\end{align}
Combining \cref{eq:backward-V} with the master equation $\partial_t\bm p=W\bm p$ cancels the terms containing $W_{mn}V_m$ and yields
\begin{align}
    \partial_tv(t,T)
    &=-\sum_e W_e(t)p_{s(e)}(t)
    \varphi_e(t,T)\varphi'_e(t,T)\nonumber\\
    &=-\sum_e j_e(t)\varphi_e(t,T)\varphi'_e(t,T)\,.
\end{align}
Since $v(T,T)=0$,
\begin{align}
    \label{eq:gf-dynamical-term}
    v(0,T)=\int_0^Tdt\sum_e
    j_e(t)\varphi_e(t,T)\varphi'_e(t,T)\,.
\end{align}
Here $j_e(t)\equiv W_e(t)p_{s(e)}(t)$ is the directed probability flux.
Finally, inserting \cref{eq:gf-dynamical-term} into \cref{eq:gf-cov} gives
\begin{align}
\label{eq:DFRR-deriviation-result}
    \langle\!\langle \mathcal{Q},\mathcal{Q}'\rangle\!\rangle
    ={}&\langle\!\langle \mathcal{Q},\mathcal{Q}'\rangle\!\rangle_0
    +\frac{1}{T}\int_0^Tdt\sum_e
    j_e(t)\varphi_e(t,T)\varphi'_e(t,T)\,,
\end{align}
which is the dynamical FRR.
\section{Autonomous relaxation: spectral representation and slow-mode approximation}
\label{sec:auto}

Here we give the details of the spectral calculation for autonomous
relaxation and show how the overlap between an observable and the slow
modes controls the initial variability.  We then evaluate the
occupation-time and current observables in the three-state example of
the main text.

\subsection{Spectral representation of the initial variability }

Consider an irreducible autonomous Markov jump process with rate matrix
$W$ and a time-independent integrated observable.  We denote its local
drift by
\begin{align}
    \bm b
    \equiv \bm\nabla_p\dot Q
    =\bm o+\sum_e x_eW_e\bm e_{s(e)},
\end{align}
where $\bm e_n$ is the unit vector associated with state $n$.  The backward
Kolmogorov equation with terminal condition $\bm Q(T,T)=0$ gives
\begin{align}
\label{eq:auto-relax-backward-solution}
    \bm Q(t,T)=\int_t^T ds\,e^{W^\intercal(s-t)}\bm b\,.
\end{align}

Let $\bm r^{(k)}$ and $\bm\ell^{(k)}$ be right and left eigenvectors of
$W$, respectively,
\begin{align}
    W\bm r^{(k)}=\lambda_k\bm r^{(k)},
    \qquad
    W^\intercal\bm\ell^{(k)}=\lambda_k\bm\ell^{(k)},
\end{align}
normalized such that
$\bm r^{(k)}\cdot\bm\ell^{(k')}=\delta_{kk'}$.  The stationary mode is
$\bm r^{(0)}=\bm\pi$, $\bm\ell^{(0)}=\bm 1$, and $\lambda_0=0$.  Assuming
for simplicity that $W$ is diagonalizable,
\begin{align}
\label{eq:auto-relax-propagator}
    e^{W^\intercal s}
    =\bm1\bm\pi^\intercal
    +\sum_{k=1}^{N-1}e^{\lambda_ks}
      \bm\ell^{(k)}\bm r^{(k)\intercal}.
\end{align}
Inserting \cref{eq:auto-relax-propagator} into
\cref{eq:auto-relax-backward-solution} yields
\begin{align}
\label{eq:auto-relax-conditional-mean}
    \bm Q(t,T)={}&(T-t)(\bm\pi\cdot\bm b)\bm1+\sum_{k=1}^{N-1}\bm\ell^{(k)}
    (\bm r^{(k)}\cdot\bm b)
    \frac{e^{\lambda_k(T-t)}-1}{\lambda_k}\,.
\end{align}
The mean reads
\begin{align}
\label{eq:auto-relax-mean}
    Q(t, T) = \bm p^\intercal(t) \bm Q(t, T) = (T-t)(\bm\pi\cdot\bm b)+\sum_{k=1}^{N-1}e^{\lambda_kt}(\bm p(0)\cdot\bm\ell^{(k)})
    (\bm r^{(k)}\cdot\bm b)
    \frac{e^{\lambda_k(T-t)}-1}{\lambda_k}\,.
\end{align}
The stationary contribution is independent of the initial state and
therefore cancels from the excess conditional mean.  At $t=0$ one obtains
\begin{align}
\label{eq:auto-relax-excess}
    \Delta Q_n(0,T)
    =\sum_{k=1}^{N-1}\Delta\ell_n^{(k)}
    (\bm r^{(k)}\cdot\bm b)
    \frac{e^{\lambda_kT}-1}{\lambda_k}\,,
\end{align}
where
$\Delta\ell_n^{(k)}\equiv\ell_n^{(k)}
-\bm p(0)\cdot\bm\ell^{(k)}$.  

\textit{Slow mode approximation---}For now suppose that a single real mode $\lambda_1=-\gamma$ is
slower than all other modes, the observable overlaps this mode, and the
initial ensemble has nonzero variance along it.  Keeping this mode in
\cref{eq:auto-relax-conditional-mean,eq:auto-relax-mean}, we find
\begin{align}
\label{eq:slow-varphi}
    \varphi_e(t,T)
    &=x_e+\bm\Delta_e\cdot\bm Q(t,T)
    \simeq x_e+(\bm\Delta_e\cdot\bm\ell^{(1)})
    b_1f(T-t),
    \nonumber\\
    \Delta Q_n(t,T)
    &\simeq \Delta\ell_n^{(1)}(t)b_1f(T-t),
\end{align}
where $f(t)\equiv(1-e^{-\gamma t})/\gamma$,
$\Delta\ell_n^{(1)}(t)\equiv\ell_n^{(1)}
-e^{-\gamma t}\bm p(0)\cdot\bm\ell^{(1)}$,
$\bm\Delta_e\cdot\bm\ell^{(1)}
=\ell^{(1)}_{t(e)}-\ell^{(1)}_{s(e)}$, and
$b_1\equiv\bm r^{(1)}\cdot\bm b$.

Define $\operatorname{Var}_{0}(\bm\ell^{(1)})
    \equiv\sum_n p_n(0)
    [\ell_n^{(1)}-\bm p(0)\cdot\bm\ell^{(1)}]^2$ and $A_Q(\gamma)
    \equiv b_1^2\operatorname{Var}_{0}(\bm\ell^{(1)})$.
The slow-mode approximation gives
\begin{align}
\label{eq:auto-relax-slow-mode}
    \llangle\mathcal Q^2\rrangle_0
    =\frac{1}{T}\sum_n p_n(0) [\Delta Q_n(0,T)]^2
    \simeq\frac{A_Q(\gamma)}{T}f^2(T).
\end{align}
As the gap closes, let
$A_Q(\gamma)\sim\gamma^\alpha$ and assume that the slow mode remains
dominant in \cref{eq:auto-relax-excess}.  For $\gamma T\ll1$,
\begin{align}
    \llangle\mathcal Q^2\rrangle_0
    \simeq A_Q(\gamma)T
    \sim\gamma^\alpha T.
\end{align}
For fixed $\gamma$ and $\gamma T\gg1$, it instead decays as
$A_Q(\gamma)/(T\gamma^2)$.  In the slow-relaxation scaling limit
$T=u/\gamma$, with $u$ fixed,
\begin{align}
\label{eq:auto-relax-scaling}
    \llangle\mathcal Q^2\rrangle_0
    \simeq
    \frac{A_Q(\gamma)}{\gamma}
    \frac{(1-e^{-u})^2}{u}
    \sim\gamma^{\alpha-1}.
\end{align}
Thus the preparation-memory contribution is of order $1/\gamma$ over
the slowest relaxation time when the slow-mode prefactor approaches a
nonzero constant, $\alpha=0$.

\textit{Slow mode with nonzero frequency---}Let the slowest relaxation be a complex-conjugate pair $\lambda_{1,2}=\lambda_\pm=-\gamma\pm i\omega$. It describes a relaxation mode whose memory decays over the timescale $1/\gamma$ while oscillating with angular frequency $\omega$.
Introducing $z_n\equiv\Delta\ell_n^{(+)}\bigl(\bm r^{(+)}\!\cdot\bm b\bigr)$ and $F_{\gamma,\omega}(T)\equiv[1-e^{(-\gamma+i\omega)T}]/(\gamma-i\omega)$, we approximate \cref{eq:auto-relax-excess} by
\begin{align}
    \Delta Q_n(0,T)
    \simeq z_nF_{\gamma,\omega}(T)+z_n^*F_{\gamma,\omega}^*(T)
    =2\operatorname{Re}[z_nF_{\gamma,\omega}(T)]
    =2|z_n||F_{\gamma,\omega}(T)|
    \cos[\arg z_n+\arg F_{\gamma,\omega}(T)]\,.
\end{align}
Under these assumptions, the initial variability reads
\begin{align}
    \llangle\mathcal Q^2\rrangle_0
    \simeq\frac{S_Q(\gamma,\omega,T)}{T}
    \frac{1-2e^{-\gamma T}\cos(\omega T)+e^{-2\gamma T}}
    {\gamma^2+\omega^2}\,,
\end{align}
where $S_Q(\gamma,\omega,T)\equiv4\sum_np_n(0)|z_n|^2
\cos^2[\arg z_n+\arg F_{\gamma,\omega}(T)]
\leq A_Q^+$ and $A_Q^+\equiv4\sum_np_n(0)|z_n|^2$. 

To obtain the gap-closing scaling, set $T=u/\gamma$ and $\rho=\omega/\gamma$. 
For a fixed $N$-state Markov generator, the bound $|\omega|/\gamma\leq\cot(\pi/N)$~\cite{dmitriev1946characteristic,ohga2023thermodynamic} ensures that $\rho=\omega/\gamma=\mathcal O(1)$ as $\gamma\to0$.
The time-dependent factor then becomes $[1-2e^{-u}\cos(\rho u)+e^{-2u}]/[\gamma u(1+\rho^2)]$. Therefore, if the complex-mode prefactor scales as $\gamma^\alpha$ and remains nonzero, the initial variability scales as $\llangle\mathcal Q^2\rrangle_0\sim\gamma^{\alpha-1}$. Thus, the long-lived scaling remains unchanged. 

\subsection{Example: Three-state molecular motor}
\label{subsec:three-state-relaxation-details}

We order the states as $(\mathrm X_1,\mathrm X_2,\mathrm X_3)$ and write
$a\equiv e^{\Delta\mu/2}$.  The positive directions of reactions $1$,
$2$, and $3$ are $\mathrm X_1\to\mathrm X_3$,
$\mathrm X_1\to\mathrm X_2$, and $\mathrm X_2\to\mathrm X_3$,
respectively.  With the convention
$\partial_t\bm p=W\bm p$, the rate matrix is
\begin{align}
\label{eq:three-state-relax-generator}
    W=
    \begin{pmatrix}
        -a-\varepsilon & \varepsilon & a^{-1}\\
        \varepsilon & -2\varepsilon & \varepsilon\\
        a & \varepsilon & -a^{-1}-\varepsilon
    \end{pmatrix}.
\end{align}
Its characteristic polynomial factorizes as
\begin{align}
\label{eq:three-state-characteristic}
    \det(\lambda I-W)
    =\lambda(\lambda+\rho)(\lambda+\kappa),
    \qquad
    \rho\equiv3\varepsilon,
    \quad
    \kappa\equiv a+a^{-1}+\varepsilon.
\end{align}
The two relaxation rates are therefore $\rho$ and $\kappa$.  The
spectral gap is $\gamma=\min\{\rho,\kappa\}$.  In the weak-link regime,
$\gamma=\rho=3\varepsilon$; this remains true throughout the range
$\varepsilon\leq1$  at $\Delta\mu=1$.
We work in this regime below.

Away from the modal degeneracy $\rho=\kappa$, a convenient
biorthogonal choice for the two nonstationary modes is
\begin{align}
\label{eq:three-state-relax-eigenvectors}
    \bm\ell^{(1)}
    &=(-1,2,-1)^\intercal,
    &
    \bm r^{(1)}
    &=\frac{1}{3(\kappa-\rho)}
    \begin{pmatrix}
        \varepsilon-a^{-1}\\
        \kappa-\rho\\
        \varepsilon-a
    \end{pmatrix},
    \nonumber\\
    \bm\ell^{(2)}
    &=\frac12
    \begin{pmatrix}1\\0\\-1\end{pmatrix}
    +\frac{a-a^{-1}}{2\kappa(\kappa-\rho)}
    \begin{pmatrix}
        a+a^{-1}-\varepsilon\\
        -2\varepsilon\\
        a+a^{-1}-\varepsilon
    \end{pmatrix},
    &
    \bm r^{(2)}
    &=\begin{pmatrix}1\\0\\-1\end{pmatrix}.
\end{align}
They satisfy $W\bm r^{(k)}=\lambda_k\bm r^{(k)}$,
$W^\intercal\bm\ell^{(k)}=\lambda_k\bm\ell^{(k)}$, and
$\bm r^{(k)}\cdot\bm\ell^{(k')}=\delta_{kk'}$, with
$\lambda_1=-\rho$ and $\lambda_2=-\kappa$.

\subsubsection{Occupation time of state 2}

For the occupation time
$\mathcal O=\int_0^Tdt\,\delta_{n_t,2}$, the local drift is
$\bm b=\bm e_2$.  In this case the fast mode has zero
overlap with $\bm b$, since $\bm b\cdot\bm r^{(2)}=0$.
The slow-mode result is therefore exact, rather than asymptotic:
\begin{align}
    e^{W^\intercal s}\bm b
    =\frac13\bm1+\frac{e^{-\gamma s}}{3}\bm\ell^{(1)}.
\end{align}
Using \cref{eq:auto-relax-conditional-mean}, we find the conditional
means
\begin{align}
    \bm O(t,T)
    =\frac{T-t}{3}\bm1
    +\frac13f(T-t)\bm\ell^{(1)}.
\end{align}
For the initial preparation
$\bm p(0)=(\theta,1-\theta,0)^\intercal$, the mean and excess
conditional means are
\begin{align}
    O(0,T)
    &=\bm p(0)\cdot\bm O(0,T)
      =\frac{T}{3}+\frac{2-3\theta}{3}f(T),
    \nonumber\\
    \Delta\bm O(0,T)
    &=\bm O(0,T)-O(0,T)\bm1
      =f(T)
    \begin{pmatrix}
        -(1-\theta)\\
        \theta\\
        -(1-\theta)
    \end{pmatrix}.
\end{align}
Hence, the initial variability is exactly
\begin{align}
\label{eq:three-state-initial-variability}
    \llangle\mathcal O^2\rrangle_0
    =\frac{\theta(1-\theta)}{T}f^2(T).
\end{align}
Here $b_1=1/3$ and
$\operatorname{Var}_0(\bm\ell^{(1)})=9\theta(1-\theta)$, so the
prefactor in \cref{eq:auto-relax-slow-mode} is
$A_{O}=\theta(1-\theta)$, independent of $\gamma$ implying
$\alpha=0$.
On the interval $0\leq\theta\leq1/2$, this expression increases
monotonically and is largest at $\theta=1/2$.  For this equal mixture,
$\llangle\mathcal O^2\rrangle_0=f^2(T)/(4T)$,
which is the dotted curve in the corresponding panel of the main text.

\subsubsection{Current through the open reaction}

Consider now $\mathcal J=\int_0^T(dk_1-dk_{-1})$, with
$\bm b_J=(a,0,-a^{-1})^\intercal$.  In this case
$\bm b_J\cdot\bm r^{(2)}\neq0$.  For the initial state
$\bm p(0)=(\theta,1-\theta,0)^\intercal$, we use
$\operatorname{Var}_0(\bm\ell^{(1)})=9\theta(1-\theta)$ and
\begin{align}
    b_1
    =\bm b_J\cdot\bm r^{(1)}
    =\frac{\varepsilon(a-a^{-1})}
    {3(a+a^{-1}-2\varepsilon)}\,,
\end{align}
to obtain the slow-mode prefactor
\begin{align}
\label{eq:three-state-current-slow-prefactor}
    A_{J}(\gamma)
    &=9\theta(1-\theta)b_1^2=\frac{\theta(1-\theta)}{9}\gamma^2
    \tanh^2\!\left(\frac{\Delta\mu}{2}\right)
    +\mathcal O(\gamma^3).
\end{align}
Thus the slow-mode prefactor has exponent $\alpha=2$.  Unlike the
occupation time, the single-mode approximation does not describe the
full initial variability $\llangle\mathcal J^2\rrangle_0$.

Retaining both nonstationary modes gives the exact result.  For the fast
mode,
\begin{align}
    b_2
    &=\bm b_J\cdot\bm r^{(2)}
      =a+a^{-1},
    &
    \ell_1^{(2)}-\ell_2^{(2)}
    &=\frac{a-\varepsilon}
    {a+a^{-1}-2\varepsilon}.
\end{align}
Since the initial distribution has support only on states $1$ and $2$,
the exact two-mode expression is
\begin{align}
\label{eq:three-state-current-two-mode}
    \llangle\mathcal J^2\rrangle_0
    ={}&\frac{\theta(1-\theta)}
    {T\left(a+a^{-1}-2\varepsilon\right)^2}
    \Bigg[
        \frac{(a+a^{-1})(a-\varepsilon)}
        {a+a^{-1}+\varepsilon}
        \left(1-e^{-(a+a^{-1}+\varepsilon)T}\right)
        -\frac{a-a^{-1}}{3}
        \left(1-e^{-3\varepsilon T}\right)
    \Bigg]^2.
\end{align}
The square in \cref{eq:three-state-current-two-mode} includes the cross
term between the slow and fast modes.  At fixed $T$ and
$\varepsilon\to0$,
\begin{align}
    \llangle\mathcal J^2\rrangle_0
    =\frac{\theta(1-\theta)a^2}
    {T(a+a^{-1})^2}
    \left(1-e^{-(a+a^{-1})T}\right)^2
    +\mathcal O(\varepsilon),
\end{align}
so the full initial variability remains of order unity even though its
slow-mode contribution is of order $\varepsilon^2T$.

In the joint scaling limit $T=u/(3\varepsilon)$, with $u$ fixed,
\cref{eq:three-state-current-two-mode} gives
\begin{align}
\label{eq:three-state-current-two-mode-scaling}
    \llangle\mathcal J^2\rrangle_0
    =\frac{\theta(1-\theta)\varepsilon}{12u}
    \left[
        3+\left(1+2e^{-u}\right)
        \tanh\!\left(\frac{\Delta\mu}{2}\right)
    \right]^2
    +\mathcal O(\varepsilon^2).
\end{align}
Thus both modal contributions to the excess conditional current remain
finite on the slow relaxation timescale, while the normalization
$1/T=3\varepsilon/u$ makes the initial variability
$\mathcal O(\varepsilon)=\mathcal O(T^{-1})$.  Consequently, the current
has no $1/\gamma$ enhancement.
\section{Frequency-domain FRRs and responses}
\label{sec:frequency-domain}
Here we show two physically distinct ways to construct frequency-domain
relations from the dynamical FRR derived in \cref{sec:DFRR}.  

Throughout this appendix, we consider an autonomous irreducible Markov jump
process with time-independent generator $W$, prepared in its nonequilibrium
steady state (NESS), $\bm p(0)=\bm\pi$, where $W\bm\pi=0$.  For a local
parameter $\lambda_e$ that changes only the oriented rate $W_e$, we write
$j_e^{\mathrm{ss}}=W_e\pi_{s(e)}$ and assume
$\partial_{\lambda_e}\ln W_e\neq0$.  Since the steady fluxes and rate
sensitivities are time independent, the dynamical FRR common to both
constructions reads
\begin{align}
\label{eq:auto-FRR-window}
    \langle\!\langle \mathcal Q,\mathcal Q'\rangle\!\rangle
    &=\langle\!\langle \mathcal Q,\mathcal Q'\rangle\!\rangle_0
    +\sum_e
    \frac{1}{j_e^{\mathrm{ss}}
    [\partial_{\lambda_e}\ln W_e]^2}
    \frac{1}{T}\int_0^T d\tau\,
    R_{\lambda_e}(\tau,T)R'_{\lambda_e}(\tau,T).
\end{align}
For the complex harmonic observables used below, we employ the bilinear
extension of this identity.

The first formulation describes an experiment observed for a finite time.
By Fourier transform with discrete frequencies
$\omega_k=2\pi k/T$, we separate its response into slow and fast oscillations.  The zero-frequency
part corresponds to a steady perturbation, while the other frequencies probe
time-dependent perturbations.  At finite observation time, the initial-variability
term can remain nonzero even for the NESS preparation.

The second formulation describes a stationary system probed for a long
time at a chosen frequency.  In the long-time scaling, the initial-variability
contribution vanishes, and we obtain a direct relation between spontaneous
fluctuations at that frequency and the response to an oscillating perturbation.

The two formulations are complementary: the first resolves a finite-time
response by frequency, whereas the second gives the stationary power
spectrum.  Both follow from the same dynamical FRR.

\subsection{Fourier decomposition of response kernels}
\label{subsec:finite-window-response-spectrum}

For a real integrated observable $\mathcal Q$, setting
$\mathcal Q'=\mathcal Q$ in \cref{eq:auto-FRR-window} gives its finite-window
variance FRR.

For fixed $T$, define the Fourier coefficients of the 
response kernel by
\begin{align}
\label{eq:finite-window-response-coefficients}
    \hat R_{\lambda_e}^{\,k}(T)
    \equiv
    \frac{1}{T}\int_0^T d\tau\,
    R_{\lambda_e}(\tau,T)e^{i\omega_k\tau},
    \qquad
    \omega_k\equiv\frac{2\pi k}{T}.
\end{align}
Parseval's identity turns \cref{eq:auto-FRR-window} into the exact
finite-window spectrum FRR:
\begin{align}
\label{eq:FRR-Fourier-window}
    \langle\!\langle \mathcal Q^2\rangle\!\rangle
    &=
    \langle\!\langle \mathcal Q^2\rangle\!\rangle_0
    +\sum_e
    \frac{1}{j_e^{\mathrm{ss}}
    [\partial_{\lambda_e}\ln W_e]^2}
    \sum_{k=-\infty}^{\infty}
    \left|\hat R_{\lambda_e}^{\,k}(T)\right|^2 .
\end{align}
Here, ``finite-window'' refers to the discrete Fourier decomposition
of the response kernel on $[0,T]$, in contrast to the stationary
power-spectrum FRR obtained in the long-time limit below.
We notice that the zero mode is the time-averaged response,
$\hat R_{\lambda_e}^{\,0}(T)=\overline R_{\lambda_e}(T)$, which controls
the response to a step perturbation.  Since every nonzero-mode contribution
in \cref{eq:FRR-Fourier-window} is nonnegative, retaining only $k=0$
gives
\begin{align}
\label{eq:FRI-zero-mode}
    \langle\!\langle \mathcal Q^2\rangle\!\rangle
    &\ge
    \langle\!\langle \mathcal Q^2\rangle\!\rangle_0
    +\sum_e
    \frac{\overline R_{\lambda_e}^2(T)}
    {j_e^{\mathrm{ss}}[\partial_{\lambda_e}\ln W_e]^2}.
\end{align}
Dropping the positive initial-variability term recovers the looser FRI of
\cite{kwon2024fluctuation}.  

\subsection{Stationary power-spectrum FRR}
\label{subsec:stationary-power-spectrum}

For the stationary power spectrum, take time-independent weights $\bm o$ and
$\bm x$.  We define the instantaneous
stochastic observable by
\begin{align}
    d\mathcal Q(t)
    \equiv \dot{\mathcal Q}(t)dt
    =o_{n_t}dt+\sum_e x_e\,dk_e(t).
\end{align}
For a real frequency $\omega$, its finite-window Fourier transform is the
harmonic observable
\begin{align}
\label{eq:finite-harmonic-observable}
    \mathcal Q_T^\omega
    &\equiv \int_0^T dt\,e^{-i\omega t}\dot{\mathcal Q}(t)
    =\int_0^T e^{-i\omega t}\,d\mathcal Q(t)\nonumber\\
    &{}=
    \int_0^T dt\,e^{-i\omega t}o_{n_t}
    +\int_0^T e^{-i\omega t}\sum_e x_e\,dk_e(t).
\end{align}
Thus $\mathcal Q_T^\omega$ is the trajectory-dependent Fourier amplitude of
$\dot{\mathcal Q}(t)$ on $[0,T]$, and its zero-frequency component is the
integrated observable of the main text,
$\mathcal Q_T^0=\mathcal Q[\omega_{0:T}]$.
The conditional mean obeys the backward equation
\begin{align}
\label{eq:finite-kolmogorov-omega}
    -\partial_t\bm Q^\omega(t,T)
    &=
    W^\intercal\bm Q^\omega(t,T)
    +e^{-i\omega t}\bm\nabla_p\dot Q,
    \qquad
    \bm Q^\omega(T,T)=0,\\
    \bm Q^\omega(t,T)
    &=
    e^{-i\omega t}
    \int_0^{T-t}ds\,
    e^{(W^\intercal-i\omega I)s}\bm\nabla_p\dot Q \nonumber\,,
\end{align}
where
$\partial_{p_n}\dot Q=o_n+\sum_e x_eW_e\delta_{n,s(e)}$.
For $T-t\to\infty$, the edge amplitude becomes
\begin{align}
\label{eq:finite-phi-omega}
    \varphi_e^\omega(t,T)
    &= e^{-i\omega t}\psi_{e,\omega},&
    \psi_{e,\omega}
    &\equiv x_e+\bm\Delta_e\cdot K_\omega^\intercal
    \bm\nabla_p\dot Q,&
    K_\omega&\equiv(i\omega I-W)^{-1}.
\end{align}
Since $\bm\Delta_e\cdot\bm 1=0$, the same
formula holds at $\omega=0$ with $K_0=(-W)^{\mathrm D}$, where ${}^D$ is the Drazin inverse.

Introducing $\delta\dot{\mathcal Q}(t)\equiv
\dot{\mathcal Q}(t)-\langle\dot{\mathcal Q}(t)\rangle_{\mathrm{ss}}$, the
stationary cross-spectrum between $\dot{\mathcal Q}$ and
$\dot{\mathcal Q}'$ reads
\begin{align}
\label{eq:stationary-cross-spectrum}
    C_{\mathcal Q,\mathcal Q'}(\omega)
    &\equiv
    \int_{-\infty}^{\infty}d\tau\,
    e^{-i\omega\tau}
    \left\langle
    \delta\dot{\mathcal Q}(\tau)\delta\dot{\mathcal Q}'(0)
    \right\rangle_{\mathrm{ss}}=
    \lim_{T\to\infty}\frac{1}{T}
    \left\langle
    \Delta\mathcal Q_T^\omega
    \Delta\mathcal Q_T^{\prime,-\omega}
    \right\rangle .
\end{align}
Applying \cref{eq:auto-FRR-window} bilinearly to $\mathcal Q_T^\omega$ and
$\mathcal Q_T^{\prime,-\omega}$, the initial-variability term vanishes in
the stationary long-time limit, and we obtain
\begin{align}
\label{eq:finite-spectrum-dfrr}
    C_{\mathcal Q,\mathcal Q'}(\omega)
    &=
    \sum_e
    j_e^{\mathrm{ss}}\,
    \psi_{e,\omega}\psi'_{e,-\omega}.
\end{align}
Here $j_e^{\mathrm{ss}}=W_e\pi_{s(e)}$.  For rate perturbations that leave
$\bm o$ and $\bm x$ fixed, define the frequency response by
\begin{align}
\label{eq:freq-resp}
    \hat R_\lambda(\omega)
    &\equiv
    \int_0^\infty ds\,e^{-i\omega s}R_\lambda(s),&
    \delta\dot Q(t)
    &=
    \int_{-\infty}^t d\tau\,
    R_\lambda(t-\tau)\delta\lambda(\tau)
    +O[(\delta\lambda)^2],
\end{align}
where $\dot Q(t)=\langle\dot{\mathcal Q}(t)\rangle$; primed response
symbols refer to $\dot Q'$.  The response kernel of the harmonic observable
reads
\begin{align}
\label{eq:harmonic-response-kernel}
    R^{\mathcal Q^\omega}_\lambda(\tau,T)
    &=
    \int_\tau^Tdt\,e^{-i\omega t}R_\lambda(t-\tau) = 
    e^{-i\omega\tau}
    \int_0^{T-\tau}ds\,e^{-i\omega s}R_\lambda(s)
    \xrightarrow[T-\tau\to\infty]{}
    e^{-i\omega\tau}\hat R_\lambda(\omega).
\end{align}
Using \cref{eq:general-kernel-SM,eq:finite-phi-omega}, we find
\begin{align}
\label{eq:inst-resp-kernel}
    \hat R_\lambda(\omega)
    &=
    \sum_e
    \psi_{e,\omega}
    j_e^{\mathrm{ss}}\,
    \partial_\lambda\ln W_e .
\end{align}
For a local perturbation $\lambda_e$ that changes only the oriented rate
$W_e$,
\begin{align}
\label{eq:inst-local-resp-kernel}
    \hat R_{\lambda_e}(\omega)
    =
    \psi_{e,\omega}
    j_e^{\mathrm{ss}}\,
    \partial_{\lambda_e}\ln W_e .
\end{align}
For state and current observables, $x_{-e}=-x_e$, this gives
\begin{align}
\label{eq:frequency-mutual-linearity}
    \frac{\hat R_{W_e}(\omega)}
    {\hat R_{W_{-e}}(\omega)}
    =
    -\frac{\pi_{s(e)}}{\pi_{t(e)}},
\end{align}
which is the frequency-domain mutual-linearity relation and the response
property of \cite{voits2026graph}.  For nonzero local sensitivities,
inserting
\cref{eq:inst-local-resp-kernel} into \cref{eq:finite-spectrum-dfrr} gives
the oriented-transition frequency-domain FRR
\begin{align}
\label{eq:freq-domain-FRR}
    C_{\mathcal Q,\mathcal Q'}(\omega)
    &=
    \sum_e
    \frac{
    \hat R_{\lambda_e}(\omega)
    \hat R'_{\lambda_e}(-\omega)}
    {j_e^{\mathrm{ss}}
    [\partial_{\lambda_e}\ln W_e]^2}.
\end{align}
This is the oriented-transition form of the frequency-domain FRR of
\cite{kwon2026nonequilibrium}.  Algebraic estimates applied to
\cref{eq:freq-domain-FRR} yield the frequency-domain FRIs of
\cite{zheng2026thermodynamic}.

\section{Periodic driving}
\label{sec:periodic}
\subsection{Dynamical FRR for Floquet state}
\label{sec:Floquet-DFRR}

Here we derive the variance in the Floquet state. We consider periodic driving for which the rates, rate sensitivities $\partial_{\lambda_e}\ln W_e$, and bounded observable weights all have period $T_p$. Let $T_M=MT_p$ and define
\begin{align}
f(\tau,T)
\equiv\sum_e
\frac{R_{\lambda_e}^2(\tau,T)}
{j_e(\tau)[\partial_{\lambda_e}\ln W_e(\tau)]^2}\,.
\end{align}
The dynamical FRR reads
\begin{align}
\label{eq:Floquet-finite-DFRR}
\llangle\mathcal Q^2[\omega_{0:T_M}]\rrangle
=\llangle\mathcal Q^2\rrangle_0
+\frac{1}{T_M}\int_0^{T_M}d\tau\,f(\tau,T_M)\,.
\end{align}
Partitioning the full interval into periods gives
\begin{align}
\label{eq:Floquet-period-partition}
\frac{1}{T_M}\int_0^{T_M}d\tau\,f(\tau,T_M)
=\frac{1}{T_M}\sum_{k=0}^{M-1}\int_0^{T_p}ds\,
f(kT_p+s,T_M)\,.
\end{align}
For $0\leq s<T_p$ in a bulk period, $1\ll k\ll M$, we define the Floquet response as
\begin{align}
\label{eq:Floquet-bulk-response}
R_{\lambda_e}^{\mathrm F}(s)
&\equiv\lim_{\substack{M\to\infty\\1\ll k\ll M}}
R_{\lambda_e}\!\left(kT_p+s,T_M\right)\,.
\end{align}
This limit assumes that the periodically driven dynamics relaxes to a unique stable Floquet state within a finite memory time. 
In the Floquet bulk, $j_e(kT_p+s)\to j_e^{\mathrm F}(s)$, while periodicity gives $\partial_{\lambda_e}\ln W_e(kT_p+s)=\partial_{\lambda_e}\ln W_e(s)$. Therefore,
\begin{align}
f(kT_p+s,T_M)&\rightarrow 
f^{\mathrm F}(s)\equiv\sum_e
\frac{[R_{\lambda_e}^{\mathrm F}(s)]^2}
{j_e^{\mathrm F}(s)[\partial_{\lambda_e}\ln W_e(s)]^2}\,.
\end{align}
We notice that the boundary terms with $k=\mathcal{O}(1)$ and $M-k=\mathcal{O}(1)$ in \cref{eq:Floquet-period-partition} contribute only as $\mathcal{O}(M^{-1})$. 
The remaining bulk contribution is periodic and can therefore be averaged over one driving period:
\begin{align}
\label{eq:Floquet-bulk-average}
\lim_{M\to\infty}\frac{1}{T_M}
\int_0^{T_M}d\tau\,f(\tau,T_M)
=\frac{1}{T_p}\int_0^{T_p}ds\,f^{\mathrm F}(s)\,.
\end{align}
In addition we assume that $\Delta Q_n(0,T_M)$ saturate at $\mathcal O(1)$ rather than growing with $T_M$. Since the initial variability is normalized by $T_M$, it scales as
\begin{align}
\llangle\mathcal Q^2\rrangle_0 = \frac{1}{T_M}\sum_np_n(0)\Delta Q^2_n(0,T_M)
=\mathcal O(T_M^{-1})
=\mathcal O(M^{-1}).
\end{align}
Combining this result with \cref{eq:Floquet-finite-DFRR,eq:Floquet-bulk-average}, we prove Eq.~(19) from the main text  
\begin{align}
\llangle\mathcal Q^2\rrangle_{\mathrm F}
=\frac{1}{T_p}\int_0^{T_p}d\tau\sum_e
\frac{[R_{\lambda_e}^{\mathrm F}(\tau)]^2}
{j_e^{\mathrm F}(\tau)[\partial_{\lambda_e}\ln W_e(\tau)]^2}\,,
\end{align}
The essential step is to average the full interval $[0,T_M]$, rather than its terminal period.


\subsection{Driven three-state finite-time current fluctuations}
\label{sec:driven-three-state}

Here we apply the dynamical FRR to a driven
version of the three-state molecular motor discussed in the main text.  The
states are ordered as $(X_1,X_2,X_3)$, and the three reversible reactions are
\begin{align}
    \mathrm{F}+X_1 &\rightleftharpoons \mathrm{W}+X_3,
    &
    X_1 &\rightleftharpoons X_2,
    &
    X_2 &\rightleftharpoons X_3.
\end{align}
The first reaction is the open reaction coupled to the chemostatted species
$\mathrm F$ and $\mathrm W$.  Its time-dependent rates are
\begin{align}
    W_{\pm1}(t)=\exp\!\left[\pm\frac{\Delta\mu(t)}{2}\right],
    \label{eq:driven-three-state-open-rates}
\end{align}
whereas the remaining rates are
$W_{\pm2}=W_{\pm3}=\varepsilon$.  All simulations start from $\bm p(0)=(1/2,1/2,0)^\intercal$.
For fixed affinity, the two nonzero instantaneous relaxation rates are
$3\varepsilon$ and $2\cosh[\Delta\mu(t)/2]+\varepsilon$ implying the spectral gap $\gamma=3\varepsilon$.
Rates and times are expressed in units where the open-reaction attempt rate
in \cref{eq:driven-three-state-open-rates} is one.  We compare
\begin{align}
    \text{slow kinetic:}\quad &
    \varepsilon_{\rm s}=0.02\,,
    &\gamma_{\rm s}=0.06\,,\quad
    &\tau_{\rm s}=\gamma_{\rm s}^{-1}=50/3,
    \nonumber\\
    \text{normal kinetic:}\quad &
    \varepsilon_{\rm n}=1/3,
    &\gamma_{\rm n}=1\,,\quad
    &\tau_{\rm n}=1.
    \label{eq:driven-three-state-relaxation-cases}
\end{align}
For each kinetic case $i\in\{\mathrm s,\mathrm n\}$, we plot time directly
as $\gamma_i t$ and final time as $\gamma_i T$, while fluctuation values defined below remain in absolute physical
units and are not divided by $\gamma_i$.

Let $dk_{\pm1}(t)$ count jumps along the forward and reverse directions of
reaction~1.  The integrated current over $[0,T]$ is
\begin{align}
    \mathcal J_1
    =\int_0^T\!\left[dk_1(t)-dk_{-1}(t)\right].
    \label{eq:driven-three-state-current}
\end{align}
The finite-time scaled variance plotted below is
\begin{align}
    \llangle\mathcal J_1^2\rrangle
    \equiv
    \frac{1}{T}
    \left(\langle\mathcal J_1^2\rangle-\langle\mathcal J_1\rangle^2\right).
    \label{eq:driven-three-state-scaled-variance}
\end{align}
To resolve its dependence on the preparation, denote by
$J_{1,n}(t,T)$ the conditional mean current accumulated over $[t,T]$ given
that the system is in state $n$ at time $t$, and set
$J_1(t,T)=\sum_n p_n(t)J_{1,n}(t,T)$ and
$\Delta J_{1,n}(t,T)=J_{1,n}(t,T)-J_1(t,T)$.  The initial variability reads
\begin{align}
    \llangle\mathcal J_1^2\rrangle_0
    =
    \frac{1}{T}\sum_n p_n(0)
    \left[\Delta J_{1,n}(0,T)\right]^2.
    \label{eq:driven-three-state-initial}
\end{align}
It is the variance, over the initially sampled state, of the corresponding
conditional mean current.
Defining
\begin{align}
    \varphi_e(t,T)
    =
    x_e+J_{1,t(e)}(t,T)-J_{1,s(e)}(t,T),
\end{align}
the dynamically generated contribution is
\begin{align}
    \llangle\mathcal J_1^2\rrangle_{\mathrm{dyn}}
    =
    \frac{1}{T}\int_0^T\!dt\,
    \sum_e j_e(t)\varphi_e^2(t,T).
    \label{eq:driven-three-state-dynamical}
\end{align}
The exact decomposition therefore reads
\begin{align}
    \llangle\mathcal J_1^2\rrangle
    =
    \llangle\mathcal J_1^2\rrangle_0
    +
    \llangle\mathcal J_1^2\rrangle_{\mathrm{dyn}}.
    \label{eq:driven-three-state-decomposition}
\end{align}
The simulations evaluate the conditional means and dynamical term through
the backward finite-time moment equations and independently check their sum
against the forward moment equations.

\begin{figure}[!t]
    \centering
    \includegraphics[width=\textwidth]{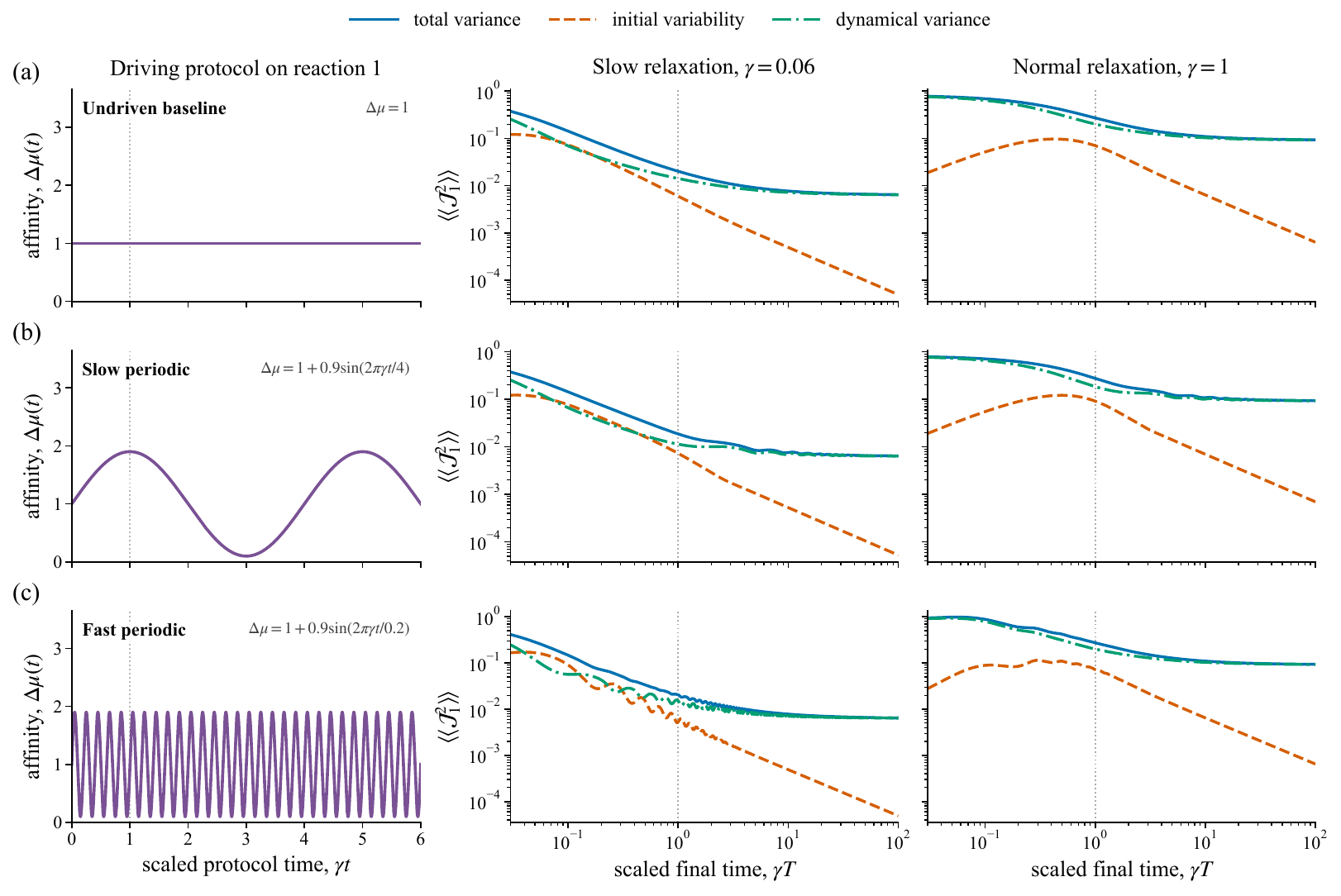}
    \caption{Finite-time variance decomposition for the driven three-state
    motor with slow and normal relaxation.  Rows (a)--(c) show the autonomous
    undriven nonequilibrium baseline with constant $\Delta\mu=1$, the sinusoidal
    protocol of scaled period $4$, and the sinusoidal protocol of scaled period
    $0.2$, respectively.  Both sinusoids have mean affinity $1$ and amplitude
    $0.9$.  In each kinetic case the axes display $\gamma_i t$ and
    $\gamma_iT$, so its physical driving periods are the displayed periods
    divided by $\gamma_i$.  The left column gives $\Delta\mu_i(t)$ versus
    $\gamma_i t$.
    The middle column gives the absolute variance decomposition
    for $\varepsilon_{\rm s}=0.02$ and $\gamma_{\rm s}=0.06$, while the right
    column gives it for $\varepsilon_{\rm n}=1/3$ and $\gamma_{\rm n}=1$.
    Blue solid, orange dashed, and green dash-dotted curves denote
    $\llangle\mathcal J_1^2\rrangle$,
    $\llangle\mathcal J_1^2\rrangle_0$, and
    $\llangle\mathcal J_1^2\rrangle_{\mathrm{dyn}}$, respectively, versus
    $\gamma_iT$.  The fluctuation values are not divided by $\gamma_i$.  Both
    variance columns extend to $\gamma_iT=100$ on a phase-resolved grid with spacing
    $0.02$ after the short logarithmic segment.  The preparation is
    $\bm p(0)=(1/2,1/2,0)^\intercal$.  Gray dotted lines in the variance
    columns mark one relaxation time, $\gamma_iT=1$.}
    \label{fig:driven-three-state-relaxation-comparison}
\end{figure}

The three rows of
\cref{fig:driven-three-state-relaxation-comparison} use the same protocol as
a function of each model's scaled time $\gamma_i t$:
\begin{align}
    \text{(a)}\quad&
    \Delta\mu_i(t)=1,
   \quad\quad
    \text{(b)}\quad
    \Delta\mu_i(t)=1+0.9\sin\!\left(\frac{2\pi\gamma_i t}{4}\right),
    \quad\quad
    \text{(c)}\quad
    \Delta\mu_i(t)=1+0.9\sin\!\left(\frac{2\pi\gamma_i t}{0.2}\right).
\end{align}
Case~(a) is an undriven \emph{nonequilibrium} baseline: the constant, nonzero
cycle affinity is $\Delta\mu=1$, so this row is neither an equilibrium
process nor a zero-affinity reference. Cases~(b) and (c) differ by the scales of the driving periods. 
The left column of the figure plots the common
protocol against $\gamma_i t$.  The middle and right columns plot all three
variance terms in
\cref{eq:driven-three-state-decomposition} for slow and normal relaxation,
respectively, against $\gamma_iT$ on logarithmic axes.  The dotted lines mark
$\gamma_i t=1$ in the protocol column and one relaxation time,
$\gamma_iT=1$, in both
variance columns.

At long observation times, once the periodically
driven dynamics has approached its Floquet state, the conditional
mean differences in \cref{eq:driven-three-state-initial} remain bounded and
the factor $1/T$ normalization makes the initial variability decay as
$\mathcal O(T^{-1})$.  The total and dynamically generated scaled variances,
by contrast, approach the same nonzero cycle-averaged plateau within each
kinetic case.  For all driven cases, the variance enters its
Floquet regime for $\gamma_iT$ of order unity.

\subsection{No-pumping theorem}
\label{sec:no-pumping}


\begin{figure}
    \centering
    \includegraphics[width=\linewidth]{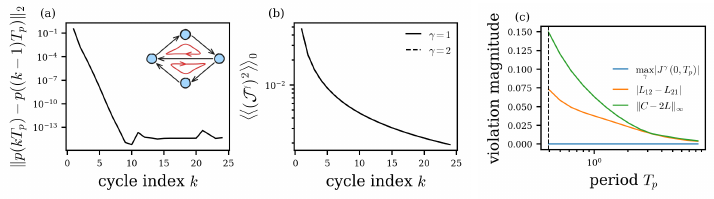}
    \caption{(a): Sketch of the two-cycle network, where the red arrows indicate the cycle currents. Starting from $\bm p(0)=(1/4,1/4,1/4,1/4)^\intercal$, the plot shows the evolution of $\|\bm p(t+T_p)-\bm p(t)\|$ under the periodic driving $E_n=E_0\cos(2\pi t/T_p-\theta_n)$ of period $T_p=\pi/8$, as a function of the cycle index $k=t/T_p$. 
    (b): The variability contribution $\langle\!\langle  (\mathcal J^\gamma)^2\rangle\!\rangle_{0}$ versus cycle index for $\gamma=1,2$ (the two curves overlap). 
    (c): In the periodic (Floquet) state, the blue curve validates the no-pumping theorem $\max_\gamma |J^\gamma(0,T_p)| = 0$, the green curve measures the lack of FDT via the norm $\|C - 2L\|_\infty$ with $C_{\gamma\gamma'}=\langle\!\langle \mathcal J^\gamma, \mathcal J^{\gamma'}\rangle\!\rangle$, and the orange curve measures nonreciprocity via $|L_{12}-L_{21}|$. 
    Both violations vanish as $T_p \to \infty$; equivalently, FDT and reciprocity are restored.
    We used $\bm B=(0,0,0,0, - 0.8)^\intercal$, $\bm\theta=(0,\pi/3,2\pi/3,\pi)$ and $E_0 = 4$.}
    \label{fig:no-pump}
\end{figure} 

We provide an alternative proof of the no-pumping theorem of Ref.~\cite{rahav2008directed}.
Consider the instantaneously detailed-balance off-diagonal rates $W_{nm}(t)=e^{B_{nm}+E_m(t)}$ with $B_{nm}=B_{mn}$. 
To avoid double counting, throughout this subsection $e>0$ denotes one arbitrarily chosen orientation of each undirected edge. Accordingly, $\Delta=(\Delta_e)_{e>0}$, $\boldsymbol{x}=(x_e)_{e>0}$. This reduced-edge convention is used only in this subsection.
In this case
$W(t)=\Delta\cdot\Gamma(t)$, with $\Gamma(t)=-D_B \cdot \Delta^\intercal \cdot D_E(t)$,
where $D_E(t)=\mathrm{diag}(\dots,e^{E_n(t)},\dots)$ and
$D_B=\mathrm{diag}(\dots,e^{B_{t(+e)s(+e)}},\dots)$.
Defining the time-independent barrier generator $W_B=-\Delta \cdot D_B \cdot \Delta^\intercal$, we write the backward Kolmogorov equation as
\begin{align}
\label{eq:no-pump-Kolmogorov}
    -\partial_t \bm{J}(t,T)
    =D_E(t)\cdot\big(W_B\cdot\bm{J}(t,T) - \Delta\cdot D_B\cdot \bm x\big)\,,
\end{align}
with $\bm{J}(T,T)=0$. We solve \cref{eq:no-pump-Kolmogorov} as $\bm{J}(t,T)=\bm c - U(t,T)\bm c$, where $\bm c$ is the constant vector satisfying $W_B\cdot\bm c - \Delta\cdot D_B\cdot \bm x = 0$ (it is defined up to an additive constant) and where $U(t,T)$ is the backward propagator generated by $W^\intercal(t)$ with $U(T,T) = 1$.
Therefore
\begin{align}
    J(\tau,\tau+T_p) &\equiv \bm p(\tau) \cdot \bm{J}(\tau,T_p+\tau)=[\bm p(\tau) - \bm p(\tau + T_p)]\cdot\bm{c}.
\end{align}
For a cyclic protocol, $E_n(\tau)=E_n(\tau + T_p)$, and in the associated Floquet periodic state, one has $\bm p(\tau + T_p)=\bm p(\tau)$, yielding
$J(\tau,\tau + T_p)=0$ and thus proving the no-pumping theorem.

\blue{One may therefore wonder whether the absence of pumped current, like that near equilibrium, implies that the response kernel satisfies Onsager reciprocity and FDT. In \cref{fig:no-pump}, we show that it is not the case using FRR for the periodically driven two-cycle network. 
Onsager symmetry and FDT are only recovered in the slow-driving limit.}
\blue{Vanishing period-integrated currents does not imply equilibrium
dynamics. Although the rates satisfy detailed balance at each fixed time,
$\bm p(t)$ generally differs from the corresponding instantaneous
equilibrium distribution during the driving. Consequently, the FDT and
Onsager reciprocity need not hold.}

\section{Connection to Fluctuation Dissipation Theorems}
\subsection{Kubo's FDT and Onsager's reciprocity}
\label{sec:KuboOnsager}

We restrict here to current observables.
Let the unperturbed process be prepared in a time-independent equilibrium distribution $\bm\pi$, so that $\bm p(0)=\bm\pi$, and suppose that, for all $t$,
\begin{align}
    W(t)\bm\pi=0\,,\qquad
    W_e(t)\pi_{s(e)}=W_{-e}(t)\pi_{t(e)}\,.
    \label{eq:Kubo-detailed-balance}
\end{align}
Hence $\bm p(t)=\bm\pi$ and the equilibrium directed fluxes satisfy
$j_e(t)=W_e(t)\pi_{s(e)}=j_{-e}(t)$.
We write $\Pi\equiv\operatorname{diag}(\pi_1,\ldots,\pi_N)$.

For the force decomposition $S_e(t)=\sum_\gamma X_{e\gamma}(t)F_\gamma(t)$ introduced in the main text,
$\partial_{F_{\gamma'}(t)}\ln W_e(t)=X_{e\gamma'}(t)/2$, evaluated at the detailed-balance reference $\bm F=0$.
The current conjugate to $F_\gamma$ is
\begin{align}
    \mathcal J^\gamma[\omega_{0:T}]
    =\int_0^Tdt\sum_e X_{e\gamma}(t)\,dk_e(t)\,.
    \label{eq:Kubo-current}
\end{align}
Let $\bm J^\gamma(t,T)$ be its vector of conditional means and, following the main-text oriented-edge definition, set
\begin{align}
    \varphi_e^\gamma(t,T)
    \equiv X_{e\gamma}(t)+\bm\Delta_e\cdot\bm J^\gamma(t,T)\,.
    \label{eq:Kubo-varphi}
\end{align}
Both $X_{e\gamma}$ and $\varphi_e^\gamma$ are antisymmetric under $e\mapsto-e$.
The integrated response of the mean current $J^\gamma(0,T)$ to $F_{\gamma'}$ is therefore
\begin{align}
    L_{\gamma\gamma'}(T)
    &\equiv\frac{1}{T}\int_0^Tdt\,
    \frac{\delta J^\gamma(0,T)}{\delta F_{\gamma'}(t)} = \frac{1}{2T}\int_0^Tdt\sum_e
    j_e(t)\varphi_e^\gamma(t,T)X_{e\gamma'}(t)\,.
    \label{eq:Kubo-L}
\end{align}

We first allow $W(t)$ and $X_{e\gamma}(t)$ to depend on time while the equilibrium distribution remains fixed.
For $\alpha\in\{\gamma,\gamma'\}$, introduce
\begin{align}
    f_n^\alpha(t)
    \equiv\sum_e\delta_{n,s(e)}W_e(t)X_{e\alpha}(t)\,.
    \label{eq:Kubo-f}
\end{align}
The backward Kolmogorov equation for the conditional current becomes
$-\partial_t\bm J^\alpha=W^\intercal\bm J^\alpha+\bm f^\alpha$, with
$\bm J^\alpha(T,T)=0$.
Detailed balance and the oriented-edge pairing give, for arbitrary state vectors $\bm u$ and $\bm v$,
\begin{align}
    \bm u\cdot\Pi W^\intercal(t)\cdot\bm v
    &=-\frac{1}{2}\sum_e j_e(t)
    (\bm\Delta_e\cdot\bm u)(\bm\Delta_e\cdot\bm v)\,,
    \label{eq:Kubo-DB-W}\\
    \bm u\cdot\Pi\cdot\bm f^\alpha(t)
    &=-\frac{1}{2}\sum_e j_e(t)X_{e\alpha}(t)
    \bm\Delta_e\cdot\bm u\,.
    \label{eq:Kubo-DB-f}
\end{align}
Here the factors $1/2$ compensate for summing over both $e$ and $-e$.
Using these identities in the two backward equations yields
\begin{align}
    \partial_t\big[\bm J^\gamma\cdot\Pi\cdot\bm J^{\gamma'}\big]
    =\frac{1}{2}\sum_e j_e(t)\big[
    2\varphi_e^\gamma\varphi_e^{\gamma'}
    -\varphi_e^{\gamma'}X_{e\gamma}
    -\varphi_e^\gamma X_{e\gamma'}\big]\,.
    \label{eq:Kubo-ode}
\end{align}
The time arguments on the $\bm J$ and $\varphi$ factors are implicit in this equation.

The equilibrium mean of each current in \cref{eq:Kubo-current} vanishes by pairwise cancelation of $e$ and $-e$.
Consequently, its initial-variability contribution is
$\langle\!\langle\mathcal J^\gamma,\mathcal J^{\gamma'}\rangle\!\rangle_0
=T^{-1}\bm J^\gamma(0,T)\cdot\Pi\cdot\bm J^{\gamma'}(0,T)$.
Integrating \cref{eq:Kubo-ode}, using the terminal condition and the oriented-edge FRR of the main text, gives the symmetrized response relation
\begin{align}
    \langle\!\langle\mathcal J^\gamma,\mathcal J^{\gamma'}\rangle\!\rangle
    &=\frac{1}{2T}\int_0^Tdt\sum_e j_e(t)
    \big[\varphi_e^{\gamma'}(t,T)X_{e\gamma}(t)
    +\varphi_e^\gamma(t,T)X_{e\gamma'}(t)\big] \nonumber\\
    &=L_{\gamma'\gamma}(T)+L_{\gamma\gamma'}(T)\,.
    \label{eq:Kubo-driving}
\end{align}

To obtain the standard Onsager relation, we additionally impose autonomous rates and time-independent current weights, $W(t)=W$ and $X_{e\alpha}(t)=X_{e\alpha}$.
Then $\bm f^\alpha$ is constant and the backward equation has the solution
\begin{align}
    \bm J^\alpha(t,T)
    =\int_0^{T-t}du\,e^{W^\intercal u}\bm f^\alpha\,.
    \label{eq:Kubo-backward-solution}
\end{align}
The part of \cref{eq:Kubo-L} containing $X_{e\gamma}X_{e\gamma'}$ is symmetric, while \cref{eq:Kubo-DB-f} gives
$\sum_e j_eX_{e\gamma'}\bm\Delta_e\cdot\bm J^\gamma
=-2\bm f^{\gamma'}\cdot\Pi\cdot\bm J^\gamma$.
It follows that
\begin{align}
    L_{\gamma\gamma'}(T)-L_{\gamma'\gamma}(T)
    ={}&-\frac{1}{T}\int_0^Tdt\int_0^{T-t}du\,
    \bm f^{\gamma'}\cdot
    \big(\Pi e^{W^\intercal u}-e^{Wu}\Pi\big)
    \cdot\bm f^\gamma=0\,.
    \label{eq:Onsager}
\end{align}
The last equality follows from detailed balance, $W\Pi=\Pi W^\intercal$, which implies
$e^{Wu}\Pi=\Pi e^{W^\intercal u}$.
Combining \cref{eq:Kubo-driving,eq:Onsager} yields the finite-time FDT
\begin{align}
    \langle\!\langle\mathcal J^\gamma,\mathcal J^{\gamma'}\rangle\!\rangle
    =2L_{\gamma\gamma'}(T)\,.
    \label{eq:Kubo-FDT}
\end{align}

\subsection{Covariance formula for responses}
\label{sec:discrepancy}
We use our theory to derive the response formula (3) in~\cite{zheng2025nonlinear}, which expresses the response of the mean $Q$ to a time-dependent perturbation $\lambda(t)$ in terms of a covariance with the auxiliary observable $\hat{\mathcal{Q}} \equiv \int_0^T  \big(\sum_e \hat{x}_e(\tau)dk_e(\tau) + \hat{o}_n(\tau) d\tau\big)$ with $\hat{o}_n = - \sum_e \delta_{ns(e)}\hat x_e W_e$. 
In this case, Eq.~(5) simplifies to $-\partial_t \hat{\bm{Q}}(t,T) = W^\intercal \hat{\bm{Q}}(t,T)$ with $\hat{\bm Q}(T,T) = 0$, which has a solution $\hat{Q}_n(t,T) = 0$ with $\hat \varphi_e(t,T) \equiv \hat x_e(t) + \Delta_e\cdot\bm{\hat Q} = \hat x_e(t)$.
Since $\hat{\mathcal{Q}}$ is the time-integrated discrepancy of the stochastic current from its predicted drift, its mean is zero. 
Inserting $\mathcal{Q}' = \hat{\mathcal{Q}}$ with $\hat\varphi_e(\tau,T) = \hat{x}_e(\tau) = \partial_{\lambda} \ln W_{e}(\tau)$ in \cref{eq:DFRR-deriviation-result} with vanishing the cross initial variability $\llangle \mathcal{Q}, \hat{\mathcal{Q}}\rrangle_0 = 0$ due to $\hat Q_n = 0$, we find
\begin{align}
\label{eq:resp-cov}
    \langle\!\langle \mathcal Q, \hat{\mathcal{Q}} \rangle\!\rangle = \frac{1}{T}\int_0^Td\tau \sum_e \varphi_e(\tau,T)j_{e}(\tau) \partial_{\lambda}\ln W_e(\tau) = \frac{1}{T}\int_0^Td\tau\frac{\delta Q(0, T)}{\delta \lambda(\tau)} = \overline{R}_{\lambda}(T)\,,
\end{align}
where we used $R_\lambda = \sum_e \varphi_e(\tau,T)j_{e}(\tau) \partial_{\lambda}\ln W_e(\tau)$.
\Cref{eq:resp-cov} can be derived from \cite{zheng2025nonlinear} assuming that $\bm x$ is independent of $\lambda(t)$. 
By applying \cref{eq:DFRR-deriviation-result} to $\hat{\mathcal{Q}}^{\gamma}$ and $\hat{\mathcal{Q}}^{\gamma'}$ with $\hat{x}_e=X_{e\gamma}=-X_{-e\gamma}$ and $\hat{x}'_e = X_{e\gamma'}=-X_{-e\gamma'}$, we find:
\begin{align}
\label{eq:hatQ-Kubo}
    \langle\!\langle \hat{\mathcal{Q}}^\gamma, \hat{\mathcal{Q}}^{\gamma'} \rangle\!\rangle = \frac{1}{T}\int_0^Td\tau\sum_{e>0} X_{e\gamma}(\tau) a_e(\tau) X_{e\gamma'}(\tau) = \frac{\int_0^T d\tau A_{\gamma\gamma'}(\tau)}{T} \,.
\end{align}
These observables $\hat{\mathcal{Q}}^\gamma$ satisfy the FDT for any distance from equilibrium and any driving protocols. Equivalent edge-level forms of \cref{eq:hatQ-Kubo} have appeared in recent response theories based on centered Poisson noises~\cite{kwon2025unified} or dynamical discrepancies~\cite{zheng2025nonlinear}.

\twocolumngrid
\bibliography{biblio}
\end{document}